\documentclass[preprint,showpacs,amsmath,amssymb,aps,prd,nofootinbib]{revtex4}
\usepackage{epsfig,graphicx,color,appendix}
\begin{document}

\title{\mbox{}\\[11pt]
 An extension of  tribimaximal  lepton mixing}

\author{Y. H. Ahn\footnote{Email: yhahn@phys.sinica.edu.tw},
Hai-Yang Cheng\footnote{Email: phcheng@phys.sinica.edu.tw},
and Sechul Oh\footnote{Email: scoh@phys.sinica.edu.tw}}

\affiliation{Institute of Physics, Academia Sinica, Taipei 115, Taiwan}

%\thanks{}


\date{\today}
%%%%%%%%%%%%%%%%%%%%%%%%%%%%%%%%%%%%%%%%%%%%%%%%%%%%%%%%%%%%%%%%%%%%%%%%%%%%%%
\begin{abstract}

Harrison, Perkins and Scott have proposed simple charged lepton and neutrino mass matrices that lead to
the tribimaximal mixing $U_{\rm TBM}$. We consider in this work an extension of the mass matrices so that the
leptonic mixing matrix becomes $U_{\rm PMNS}=V_L^{\ell\dagger}U_{\rm TBM}W$, where $V_L^\ell$ is a unitary
matrix needed to diagonalize the charged lepton mass matrix and $W$ measures the deviation of the neutrino
mixing matrix from the bimaximal form. Hence, corrections to $U_{\rm TBM}$ arise from both charged
lepton and neutrino sectors. Following our previous work to assume a Qin-Ma-like parametrization $V_{\rm QM}$ for the
charged lepton mixing matrix $V_L^\ell$ in which the {\it CP}-odd phase is approximately maximal,
we study the phenomenological implications in two different scenarios: $V_L^\ell=V_{\rm QM}^\dagger$ and
$V_L^\ell=V_{\rm QM}$.
We find that the latter is more
preferable, though both scenarios are consistent with the data within $3\sigma$ ranges. The predicted reactor neutrino mixing angle $\theta_{13}$ in both scenarios is consistent with the recent T2K and MINOS data.
The leptonic {\it CP} violation characterized by the Jarlskog invariant $J_{\rm CP}$ is generally of order $10^{-2}$.

\end{abstract}
\maketitle
%%%%%%%%%%%%%%%%%%%%%%%%%%%%%%%%%%%%%%%%%%%%%%%%%%%%%%%%%%%%%%%%%%%%%%%%%%%%%%%%%%%%%%%%%%%%%%%
\section{Introduction}
The large values of the solar ($\theta_{12}$) and atmospheric ($\theta_{23}$) mixing angles may be telling us about some new symmetries of leptons not presented in the quark sector and may provide a clue to the nature of the quark-lepton physics beyond the standard model. If there exists such a flavor symmetry in Nature, the tribimaximal (TBM)~\cite{HPS} pattern for the neutrino mixing will be a good zeroth order approximation to reality :
 \begin{eqnarray}
  \sin^{2}\theta_{12}=\frac{1}{3}~,\qquad\sin^{2}\theta_{23}=\frac{1}{2}~,\qquad\sin\theta_{13}=0~.
 \end{eqnarray}
For example, in a well-motivated extension of the standard model through the inclusion of $A_{4}$ discrete symmetry, the TBM pattern comes out in a natural way in the work of~\cite{He:2006dk}. Although such a flavor symmetry is realized in Nature leading to exact TBM, in general there may be some deviations from TBM.
Recent data of the T2K \cite{Abe:2011sj} and MINOS \cite{MINOS} Collaborations and the analysis based on global fits~\cite{GonzalezGarcia:2010er, Fogli:2011qn} of neutrino oscillations enter into a new phase of precise measurements of the neutrino mixing angles and mass-squared differences, indicating that the TBM mixing for three flavors of leptons should be modified. In the weak eigenstate basis, the Yukawa interactions in both neutrino and charged lepton sectors and the charged gauge interaction can be written as
 \begin{eqnarray}
 -{\cal L} &=& \frac{1}{2}\overline{\nu_{L}} ~{\cal M}_{\nu} ~(\nu_{L})^c +\overline{\ell_{L}}m_{\ell}\ell_{R}
  + \frac{g}{\sqrt{2}}W^{-}_{\mu} ~\overline{\ell_{L}}\gamma^{\mu}\nu_{L}  + {\rm H.c.} ~.
 \label{lagrangianA}
 \end{eqnarray}
When diagonalizing the neutrino and charged lepton mass matrices $U^{\dag}_{\nu}{\cal M}_{\nu}U^{\ast}_{\nu}={\rm diag}(m_{1},m_{2},m_{3}),~ U^{\dag}_{L}m_{\ell}U_{R}={\rm diag}(m_{e},m_{\mu},m_{\tau})$, one can rotate the neutrino and charged lepton fields from the weak eigenstates to the mass eigenstates $\nu_{L}\rightarrow U^{\dag}_{\nu}\nu_{L},~\ell_{L(R)}\rightarrow U^{\dag}_{L(R)}\ell_{L(R)}$.
Then we obtain the leptonic $3\times3$ unitary mixing matrix $U_{\rm PMNS}= U^{\dag}_{L}U_\nu$ from the charged current term in Eq.~(\ref{lagrangianA}).
%which reflects that, in general, the mixing matrix $V^{\ell}_{L}$ associated with the diagonalization of the charged leptons corrects a given neutrino mixing matrix and such a correction can in principle generate sizable deviations for all observables.
In the standard parametrization of the leptonic mixing matrix  $U_{\rm PMNS}$, it is expressed in terms of three mixing angles and three {\it CP}-odd phases (one for the Dirac neutrino and two for the Majorana neutrino) \cite{PDG}
 \begin{eqnarray}
  U_{\rm PMNS}={\left(\begin{array}{ccc}
   c_{13}c_{12} & c_{13}s_{12} & s_{13}e^{-i\delta_{CP}} \\
   -c_{23}s_{12}-s_{23}c_{12}s_{13}e^{i\delta_{CP}} & c_{23}c_{12}-s_{23}s_{12}s_{13}e^{i\delta_{CP}} & s_{23}c_{13}  \\
   s_{23}s_{12}-c_{23}c_{12}s_{13}e^{i\delta_{CP}} & -s_{23}c_{12}-c_{23}s_{12}s_{13}e^{i\delta_{CP}} & c_{23}c_{13}
   \end{array}\right)}P_{\nu}~,
 \label{PMNS}
 \end{eqnarray}
where $s_{ij}\equiv \sin\theta_{ij}$ and $c_{ij}\equiv \cos\theta_{ij}$, and $P_{\nu}={\rm diag}(e^{i\delta_{1}},e^{i\delta_{2}},1)$ is a diagonal phase matrix which contains two {\it CP}-violating Majorana phases, one (or a combination) of which can be in principle explored through the neutrinoless double beta ($0\nu2\beta$) decay~\cite{Schechter:1981bd}.
For the global fits of the available data from neutrino oscillation experiments, we quote two recent analyses: one by Gonzalez-Garcia {\it et al.} ~\cite{GonzalezGarcia:2010er}
 \begin{eqnarray}
  \sin^{2}\theta_{12} &=& 0.319^{+0.016~(+0.053)}_{-0.016~(-0.046)}\ , \quad\quad
  \sin\theta_{13}=0.097^{+0.052}_{-0.050}~(\leq0.217)\ , \nonumber\\
  \sin^{2}\theta_{23} &=& 0.462^{+0.082~(+0.185)}_{-0.050~(-0.124)} \ ,
 \label{exp00}
 \end{eqnarray}
in $1\sigma$ ($3\sigma$) ranges, or equivalently
 \begin{eqnarray}
  \theta_{12}=34.4^{\circ+1.0^{\circ}~(+3.2^{\circ})}_{~-1.0^{\circ}~(-2.9^{\circ})}~,
  ~~~~~\theta_{23}=42.8^{\circ+4.7^{\circ}~(+10.7^{\circ})}_{~-2.9^{\circ}~(~-7.3^{\circ})}~,
  ~~~~~\theta_{13}=5.6^{\circ+3.0^{\circ}~(+6.9^{\circ})}_{~-2.9^{\circ}~(-5.6^{\circ})}~,
 \label{exp0}
 \end{eqnarray}
and the other given by Fogli {\it et al.} with new reactor neutrino fluxes ~\cite{Fogli:2011qn}:
 \begin{eqnarray}
  \sin^{2}\theta_{12} &=& 0.312^{+0.017~(+0.052)}_{-0.006~(-0.047)}\ , \quad\quad
  \sin^2\theta_{13}=0.025^{+0.007~(+0.025)}_{-0.007~(-0.020)}\ , \nonumber \\
  \sin^{2}\theta_{23}&=&0.42^{+0.08~(+0.22)}_{-0.03~(-0.08)} \ ,
 \label{exp}
 \end{eqnarray}
corresponding to
 \begin{eqnarray}
  \theta_{12}=34.0^{\circ+1.0^{\circ}~(+3.2^{\circ})}_{~-1.0^{\circ}~(-3.0^{\circ})}~,
  ~~~~~\theta_{23}=40.4^{\circ+4.6^{\circ}~(+12.7^{\circ})}_{~-1.3^{\circ}~(~-4.7^{\circ})}~,
  ~~~~~\theta_{13}=9.1^{\circ+1.2^{\circ}~(+3.8^{\circ})}_{~-1.4^{\circ}~(-5.0^{\circ})}~.
 \label{exp1}
 \end{eqnarray}
The analysis by Fogli {\it et al.} includes the T2K \cite{Abe:2011sj} and MINOS \cite{MINOS} results. The T2K Collaboration~\cite{Abe:2011sj} has announced that the value of $\theta_{13}$ is non-zero at $90\%$ C.L. with the ranges
 \begin{eqnarray}
 0.03~(0.04)\leq\sin^{2}2\theta_{13}\leq0.28~(0.34) \ ,
 \label{T2K}
 \end{eqnarray}
or
 \begin{eqnarray}
  4.99^{\circ}~(5.77^{\circ})\leq\theta_{13}\leq15.97^{\circ}~(17.83^{\circ})~
 \label{T2K1}
 \end{eqnarray}
for $\delta_{CP}=0$, $\sin^{2}2\theta_{23}=1$ and the normal (inverted) neutrino mass hierarchy. The MINOS Collaboration found
 \begin{eqnarray}
 \sin^{2}2\theta_{13}\leq0.12~(0.20) \ ,
 \label{MINOS}
 \end{eqnarray}
with a best fit of
 \begin{eqnarray}
 \sin^{2}2\theta_{13}=0.041^{+0.047}_{-0.031}~(0.079^{+0.071}_{-0.053}) \ ,
 \label{MINOS}
 \end{eqnarray}
for $\delta_{CP}=0$, $\sin^{2}2\theta_{23}=1$ and the normal (inverted) neutrino mass hierarchy.
The experimental result of non-zero $|U_{e3}|\equiv\sin\theta_{13}$ implies that the TBM pattern should be modified.
However, properties related to the leptonic {\it CP} violation remain completely unknown yet.

%%%%%%%%%%%%%%%%%%%%%%%%%%%%%%%%%%%%%%%%%%%%%%%%%%%%%%%%%%%%%%%%%%%%%%%%%%%%%%%%%%%%%%%%%%%%%%%
The trimaximal neutrino mixing was first proposed by Cabibbo~\cite{Cabibbo:1977nk}\footnote{The matrix
originally given by Cabibbo was in the form
 \begin{eqnarray}
 V_{C}=\frac{1}{\sqrt{3}}{\left(\begin{array}{ccc}
 1 &  1 &  1 \\
 1 &  \omega &  \omega^* \\
 1 &  \omega^* &  \omega
 \end{array}\right)}~. \nonumber
 \end{eqnarray}
If one considers $A_{4}$ discrete symmetry, it will have two subgroups, namely, $Z_{2}$ and $Z_{3}$. The trimaximal matrix given in Eq. (\ref{Cabibbo}) is obtained under $Z_{3}$.}
(see also \cite{trimax})
 \begin{eqnarray}
 V_{C}=\frac{1}{\sqrt{3}}{\left(\begin{array}{ccc}
 1 &  \omega^{2} &  \omega \\
 1 &  1 &  1 \\
 1 &  \omega &  \omega^{2}
 \end{array}\right)}~,
 \label{Cabibbo}
 \end{eqnarray}
with $\omega=e^{i2\pi/3}$ being a complex cube-root of unity. This mixing matrix has maximal {\it CP}
violation with the Jarlskog invariant $|J_{ CP}|=1/(6\sqrt{3})$. However, this trimaximal mixing pattern
has been ruled out by current experimental data on neutrino oscillations. In their original work,
Harrison, Perkins and Scott (HPS)~\cite{HPS} proposed to consider the simple mass matrices
 \begin{eqnarray}
 M^{2}_{\ell}={\left(\begin{array}{ccc}
 a &  b &  b^{\ast} \\
 b^{\ast} & a &  b \\
 b &  b^{\ast}  &  a
 \end{array}\right)}~,\quad
 M^{2}_{\nu}={\left(\begin{array}{ccc}
 x & 0 & y \\
 0 & z & 0  \\
 y & 0 & x
 \end{array}\right)}~,
 \label{mass1}
 \end{eqnarray}
that can lead to the tribimaximal mixing, where $a,x,y$ and $z$ are real parameters,\footnote{ Different
from the choice of HPS, the matrix element $y$ in Eq.~(\ref{mass1}) can be in general introduced as
complex: e.g., $(M^{2}_{\nu})_{13}= y$ and $(M^{2}_{\nu})_{31} = y^*$.  This case has been considered
by Xing \cite{Xing:2002sw} who pointed out that the off-diagonal terms in $U_{\rm BM}$ will acquire a phase
from the complex $y$. It has the interesting implication that a nonzero $\sin\theta_{13}$ will result from
the phase of $y$. However, the corresponding Jarlskog invariant is exactly zero and the absence of intrinsic
{\it CP} violation makes this possibility less interesting.}
$M^{2}_{\ell}\equiv m_{\ell}m^{\dag}_{\ell}$ and
$M_\nu^2\equiv {\cal M}_{\nu}{\cal M}^{\dag}_{\nu}$.  The mass matrices  are diagonalized
by the trimaximal matrix $V_{C}$ for charged lepton fields and the bimaximal matrix $U_{\rm BM}$ defined
below for neutrino fields, that is, $V_C^\dagger M_\ell^2 V_C={\rm diag}(m_e^2,m_\mu^2,m_\tau^2)$ and
$U_{\rm BM}^\dagger M_\nu^2 U_{\rm BM}={\rm diag}(m_1^2,m_2^2,m_3^2)$.  The combination of trimaximal and
bimaximal matrices leads to the so-called TBM mixing matrix:
 \begin{eqnarray}
 U_{\rm TBM}=V^{\dag}_{C}~U_{\rm BM}={\left(\begin{array}{ccc}
 \sqrt{\frac{2}{3}} &  \frac{1}{\sqrt{3}} &  0 \\
 -\frac{1}{\sqrt{6}} &  \frac{1}{\sqrt{3}} &  -\frac{i}{\sqrt{2}} \\
 -\frac{1}{\sqrt{6}} &  \frac{1}{\sqrt{3}} &  \frac{i}{\sqrt{2}}
 \end{array}\right)}~\qquad {\rm with}~U_{\rm BM}={\left(\begin{array}{ccc}
 \frac{1}{\sqrt{2}} &  0 &  -\frac{1}{\sqrt{2}} \\
 0 &  1 &  0 \\
 \frac{1}{\sqrt{2}} &  0 &  \frac{1}{\sqrt{2}}
 \end{array}\right)}~.
 \label{HPS}
 \end{eqnarray}
It is clear by now that the tribimaximal mixing is not consistent with the recent experimental data on
the reactor mixing angle $\theta_{13}$ because of the vanishing matrix element $U_{e3}$ in $U_{\rm TBM}$.

In this work we consider an extension of the tribimaximal mixing by considering small perturbations
to the mass matrices $M_\ell^2$ and $M_\nu^2$ which we will call $M_\ell^{\prime~ 2}$ and $M_\nu^{\prime~ 2}$,
respectively (see Eq. (\ref{mass2}) below) so that $U_\nu=U_{\rm BM}W$
is no longer in the bimaximal form and $U_L = V_C V_L^\ell$ deviates from the trimaximal structure,
where $V_L^\ell$ is the unitary matrix needed to diagonalize the matrix $V_C^\dagger M_\ell^{\prime ~2} V_C$.
As a consequence, $U_{\rm PMNS}=U_L^{\dagger} U_\nu=V_L^{\ell\dagger}U_{\rm TBM}W=U_{\rm TBM}+$ small
perturbations. Hence, the corrections to the TBM pattern arise from both charged lepton and neutrino
sectors. Inspired by the T2K and MINOS measurements of a sizable reactor angle $\theta_{13}$, there exist in the
literature intensive studies of possible deviations from the exact TBM pattern. However, most of these
investigations were focused on the modification of TBM arising from either the neutrino sector~\cite{TBMnu}
or the charged lepton part~\cite{Ahn:2011ep,TBMlep}, but not both simultaneously.

The paper is organized as follows. In Sec.~II, we set up the model by making a general extension
to the charged lepton and neutrino mass matrices. Then in Sec.~III we study the phenomenological
implications by considering two different scenarios for the charged lepton mixing matrix. Our conclusions are
summarized in Sec.~IV.

%%%%%%%%%%%%%%%%%%%%%%%%%%%%%%%%%%%%%%%%%%%%%%%%%%%%%%%%%%%%%%%%%%%%%%%%%%%%%%%%%%%%%%%%%%%%%%%
\section{A simple and realistic extension}
In order to discuss the deviation from the TBM mixing, let us consider a simple and general extension of the original
proposal by HPS given in Eq.~(\ref{mass1}), by taking into account perturbative effects on the mass
matrices $M^2_{\ell}$ and $M^2_{\nu}$.
The generalized mass matrices $M^{\prime ~2}_f$ and $M^{\prime ~2}_{\nu}$ can be introduced as~\footnote{TBM could be obtained in models with different discrete symmetries, such as $S_{3},A_{4},S_{4},A_{5}$, dihedral groups, $\cdots$, etc. By considering higher order and radiative effects, the matrices in Eq.~(\ref{mass2}) can be realized. For example, we have shown in Ref.~\cite{Ahn:2011yj} that these matrices can be obtained by introducing dimension-5 operators to the Lagrangians.}
 \begin{eqnarray}
 M^{\prime ~2}_f = {\left(\begin{array}{ccc}
 a+g_{3} &  b+\chi_{3} &  b^{\ast}+\chi^{\ast}_{2} \\
 b^{\ast}+\chi^{\ast}_{3} & a+g_{2} &  b+\chi_{1} \\
 b+\chi_{2} &  b^{\ast}+\chi^{\ast}_{1} &  a+g_{1}
 \end{array}\right)}~,\quad
 M^{\prime ~2}_{\nu} = m^{2}_{0}{\left(\begin{array}{ccc}
 x' & 0 & y' \\
 0 & 1 & 0  \\
 y' & 0 & x' + \rho
 \end{array}\right)}~,
 \label{mass2}
 \end{eqnarray}
where $M^{\prime ~2}_f$ and $M^{\prime ~2}_{\nu}$ are defined as the hermitian square of the mass
matrices $M^{\prime ~2}_f \equiv m'_f m'^{\dagger}_f$ and
$M^{\prime ~2}_{\nu} \equiv {\cal M}'_{\nu} {\cal M}'^{\dagger}_{\nu}$, respectively, with the subscript
$f$ denoting charged fermion fields (charged leptons or quarks). Due to the hermiticity of $M^{\prime ~2}_f$
and $M^{\prime ~2}_{\nu}$, the parameters $a, g_{1,2,3}, m^2_0, x', y', \rho$ are real, while
$b$ and $\chi_{1,2,3}$ are complex. The parameters $g_{1,2,3}$, $\chi_{1,2,3}$ and $\rho$ represent small
perturbations. Note that the (11), (13), (22) elements ({\it i.e.,} $m_0^2 x'$, $m_0^2 y'$ and $m_0^2$)
in $M^{\prime ~2}_{\nu}$ are assumed to contain any perturbative effects on the elements $x$, $y$, and $z$
in $M^2_{\nu}$, respectively.
For simplicity, it is assumed that $y'$ is real just as the other elements in $M^{\prime ~2}_{\nu}$ and
the vanishing off-diagonal elements in $M^2_{\nu}$ remain zeros in $M^{\prime ~2}_{\nu}$.

The parameters $a$ and $b$ are encoded in~\cite{HPS} as
 \begin{eqnarray}
  a=\frac{\tilde{m}^{2}_{f_{1}}}{3}+\frac{\tilde{m}^{2}_{f_{2}}}{3}+\frac{\tilde{m}^{2}_{f_{3}}}{3}~,
  \qquad~~\quad
  b=\frac{\tilde{m}^{2}_{f_{1}}}{3}+\frac{\tilde{m}^{2}_{f_{2}}\omega^{2}}{3}+\frac{\tilde{m}^{2}_{f_{3}}\omega}{3}~,
 \end{eqnarray}
where the subscript $f_{i}$ indicates a generation of charged fermion field, and $\tilde{m}_{f_{i}}$
represents a bare mass of $f_{i}$, for example,
$\tilde{m}_{f_{1}}=\tilde{m}_{e}\ll \tilde{m}_{f_{2}}=\tilde{m}_{\mu}\ll \tilde{m}_{f_{3}}=\tilde{m}_{\tau}$
for charged lepton fields.

We first discuss the hermitian square of the neutrino mass matrix, $M^{\prime ~2}_{\nu}$,
in Eq.~(\ref{mass2}). It can be diagonalized by
 \begin{eqnarray}
 U_{\nu}={\left(\begin{array}{ccc}
 \cos\theta &  0 &  -\sin\theta \\
 0 &  1 &  0 \\
 \sin\theta &  0 &  \cos\theta
 \end{array}\right)}
 P_{\nu}={\left(\begin{array}{ccc}
 {1/\sqrt{2}} &  0 &  -{1/\sqrt{2}}\\
 0 &  1 &  0 \\
 {1/\sqrt{2}} &  0 &  {1/\sqrt{2}}
 \end{array}\right)}W \ ,
 \label{Unu}
 \end{eqnarray}
with
 \begin{eqnarray}
 \tan2\theta=-\frac{2y'}{\rho}
 \end{eqnarray}
and
\begin{eqnarray}
  W={\left(\begin{array}{ccc}
 (\cos\theta+\sin\theta)/\sqrt{2} &  0 &  (\cos\theta-\sin\theta)/\sqrt{2} \\
 0 &  1 &  0 \\
 -(\cos\theta-\sin\theta)/\sqrt{2} &  0 &  (\cos\theta+\sin\theta)/\sqrt{2}
 \end{array}\right)}P_{\nu} \ ,
\end{eqnarray}
where the diagonal phase matrix $P_{\nu}$ contains two additional phases, which can be absorbed into the neutrino mass eigenstate fields.
For a small perturbation $|\rho|~(\ll |x'|)$,  the mixing parameter $\theta$ can be expressed in terms of
 \begin{eqnarray}
 \theta=\pi/4+\epsilon  ~~~ {\rm with} ~~ |\epsilon|\ll1 ~.
 \end{eqnarray}
$W$ is then reduced to
 \begin{eqnarray}
  W=
 {\left(\begin{array}{ccc}
 \cos\epsilon &  0 &  -\sin\epsilon \\
 0 &  1 &  0 \\
 \sin\epsilon &  0 &  \cos\epsilon
 \end{array}\right)}P_{\nu}~.
 \label{epsilon}
 \end{eqnarray}
The neutrino mass eigenvalues are obtained as
 \begin{eqnarray}
  m^{2}_{1}&=&m^{2}_{0}(x' +\rho\sin^{2}\theta +y'\sin2\theta),\quad
  m^{2}_{2}=m^{2}_{0},\quad
  m^{2}_{3}=m^{2}_{0}(x' +\rho\cos^{2}\theta -y'\sin2\theta)
 \end{eqnarray}
and their  differences are given by
 \begin{eqnarray}
  \Delta m^{2}_{21} &\equiv& m^{2}_{2}-m^{2}_{1}
   =m^{2}_{0}\left(1 -x' +\rho ~\frac{1-\sin2\epsilon}{2\sin2\epsilon}\right)~,\nonumber\\
  \Delta m^{2}_{31} &\equiv& m^{2}_{3}-m^{2}_{1} =m^2_0 ~\frac{2\rho}{\sin2\epsilon}~,
 \label{masssquare}
 \end{eqnarray}
from which we have a relation $\Delta m^{2}_{21}-\frac{1}{4}\Delta m^{2}_{31}\simeq m^{2}_{2}(1-x')$.
It is well known that the sign of $\Delta m^{2}_{21}$ is positive due to the requirement of the
Mikheyev-Smirnov-Wolfenstein resonance for solar neutrinos. The sign of $\Delta m^{2}_{31}$ depends on that of $\rho/\sin2\epsilon$: $\Delta m^{2}_{31}>0$ for the normal mass spectrum and $\Delta m^{2}_{31}<0$ for the inverted one.  The quantities $m^{2}_{1},m^{2}_{2},m^{2}_{3},\theta$ (or $\epsilon$) are determined by the four parameters $m^{2}_{0},x',y',\rho$, while the Majorana phases in Eq.~(\ref{Unu}) are hidden in the squared mass eigenvalues.

We next turn to the hermitian square of the mass matrix for charged fermions in Eq.~(\ref{mass2}).
This modified charged fermion mass matrix is no longer diagonalized by $V_{C}$
 \begin{eqnarray}
 V^{\dag}_{C} M^{\prime ~2}_f V_{C}
 = {\left(\begin{array}{ccc}
 m^{2}_{a}+\eta_{11} &  \eta_{12} &  \eta_{13} \\
 \eta^{\ast}_{12} &  m^{2}_{b}+\eta_{22} &  \eta_{23} \\
 \eta^{\ast}_{13} &  \eta^{\ast}_{23} &  m^{2}_{c}+\eta_{33}
 \end{array}\right)}~,
 \label{AA}
 \end{eqnarray}
where
 \begin{eqnarray}
  m^{2}_{a}=a+b+b^{\ast}~,\qquad m^{2}_{b}=a+b~\omega+b^{\ast}~\omega^{2}~,\qquad
  m^{2}_{c}=a+b~\omega^{2}+b^{\ast}~\omega~,
 \end{eqnarray}
corresponding to $\tilde{m}^{2}_{f_{1}},~\tilde{m}^{2}_{f_{2}},~\tilde{m}^{2}_{f_{3}}$, respectively,
and $\eta_{ij}$ is composed of the combinations of $g_{1,2,3}$ and $\chi_{1,2,3}$.  To diagonalize
$V^{\dag}_{C} M^{\prime ~2}_f V_{C} = V^{f}_{L} ~{\rm diag}(m^{2}_{f_{1}},m^{2}_{f_{2}},m^{2}_{f_{3}})~
V^{f\dag}_{L}$, we need an additional matrix $V^{f}_{L}$ which can be, in general, parametrized in terms
of three mixing angles and six phases:
 \begin{eqnarray}
 V^{f}_{L}={\left(\begin{array}{ccc}
 c_{2}c_{3} &  c_{2}s_{3}e^{i\alpha_{3}} &  s_{2}e^{i\alpha_{2}} \\
 -c_{1}s_{3}e^{-i\alpha_{3}}-s_{1}s_{2}c_{3}e^{i(\alpha_{1}-\alpha_{2})} &  c_{1}c_{3}-s_{1}s_{2}s_{3}e^{i(\alpha_{1}-\alpha_{2}+\alpha_{3})} &  s_{1}c_{2}e^{i\alpha_{1}} \\
 s_{1}s_{3}e^{-i(\alpha_{1}+\alpha_{3})}-c_{1}s_{2}c_{3}e^{-i\alpha_{2}} &  -s_{1}c_{3}e^{-i\alpha_{1}}-c_{1}s_{2}s_{3}e^{i(\alpha_{3}-\alpha_{2})} &  c_{1}c_{2}
 \end{array}\right)}P_{f}~,
 \label{Vl}
 \end{eqnarray}
where $s_{i}\equiv \sin\theta_{i}$, $c_{i}\equiv \cos\theta_{i}$ and a diagonal phase matrix
$P_{f}={\rm diag}(e^{i\xi_{1}},e^{i\xi_{2}},e^{i\xi_{3}})$ which can be rotated away by the phase
redefinition of left-charged fermion fields. The charged fermion mixing matrix now reads $U_L=V_CV_L^f$.

Finally, we arrive at the general expression for the leptonic mixing matrix
\begin{eqnarray}
U_{\rm PMNS}= U_L^{\dagger}U_\nu =V_L^{\ell\dagger}U_{\rm TBM}W \ .
\end{eqnarray}
A simple and general extension of the mass matrices given in Eq.~(\ref{mass2}) thus leads to two possible sources
of corrections to the tribimaximal mixing: $V_L^\ell$ measures the deviation of the charged lepton mixing
matrix from the trimaximal form and $W$ characterizes the departure of the neutrino mixing from the
bimaximal one. The charged lepton mass matrix in Eq.~(\ref{mass2}) or (\ref{AA}) has 12 free parameters. Three of them are replaced by the phases $\xi_{1,2,3}$ in Eq.~(\ref{Vl}) which can be eliminated by a redefinition of the physical charged lepton fields. The remaining 9 parameters can be expressed in terms of $m_{e},m_{\mu},m_{\tau},\theta_{1},\theta_{2},\theta_{3},\alpha_{1},\alpha_{2},\alpha_{3}$.
>From Eqs.~(\ref{AA}) and (\ref{Vl}) the mixing angles and phases can be expressed as
 \begin{eqnarray}
  \theta_{1}&\simeq&\frac{|\eta_{23}|}{\tilde{m}^{2}_{\tau}}~,\qquad\quad
  \theta_{2}\simeq\frac{|\eta_{13}|}{\tilde{m}^{2}_{\tau}}~,\qquad\quad
  \theta_{3}\simeq\frac{|\eta_{12}|}{\tilde{m}^{2}_{\mu}}~, \qquad
  \alpha_{1}= \arg(\eta_{23}), \nonumber \\
  \alpha_{2}&\simeq& \frac{1}{2}\arg(\eta_{23})
  +\arg(\eta_{13}), \quad \alpha_{3}\simeq\frac{1}{2}\left[\arg(\eta_{13})
  -\arg(\eta_{23})\right]+\arg(\eta_{12}) \ ,
 \end{eqnarray}
with the condition $\tilde{m}^{2}_{f_{2}}\gg\eta_{22},\eta_{11}$.
In the charged fermion sector, there is a qualitative feature that distinguishes the neutrino sector from
the charged fermion one. The mass spectrum of the charged leptons exhibits a similar hierarchical pattern
to that of the down-type quarks, unlike that of the up-type quarks which show a much stronger hierarchical
pattern. For example, in terms of the Cabbibo angle $\lambda \equiv \sin\theta_{\rm C} \approx |V_{us}|$,
the fermion masses scale as ~$(m_{e},m_{\mu}) \approx (\lambda^{5},\lambda^{2}) m_{\tau}$,
$(m_{d},m_{s}) \approx (\lambda^{4},\lambda^{2}) m_{b}$ and
$(m_{u},m_{c}) \approx (\lambda^{8},\lambda^{4}) m_{t}$.
This may lead to two implications: (i) the Cabibbo-Kobayashi-Maskawa (CKM) matrix \cite{CKM} is mainly
governed by the down-type quark mixing matrix, and (ii) the charged lepton mixing matrix is similar to
that of the down-type quark one. Therefore, we shall assume that (i) $V_{\rm CKM}=V^{d\dag}_{L}$ and
$V^{u}_{L}=\mathbf{1}$, where $V^{d}_{L}~(V^{u}_{L})$ is associated with the diagonalization of the
down-type (up-type) quark mass matrix and $\mathbf{1}$ is a $3\times3$ unit matrix, and (ii) the charged
lepton mixing matrix $V^{\ell}_{L}$ has the same structure as the CKM matrix, that is,
$V_L^{\ell\dagger}=V_{\rm CKM}$ or $V_{\rm CKM}^\dagger$.

Recently, we have proposed a  simple {\it ansatz} for the charged lepton mixing matrix $V_L^\ell$, namely, it has the Qin-Ma-like parametrization in which the {\it CP}-odd phase is approximately maximal~\cite{Ahn:2011ep}. Armed with this {\it ansatz}, we notice that the 6 parameters $\theta_{1},\theta_{2},\theta_{3},\alpha_{1},\alpha_{2},\alpha_{3}$ in $V_L^\ell$ are reduced to four independent ones $f,h,\lambda,\delta$.
It has the advantage that the TBM predictions of $\sin^2\theta_{23}=1/2$ and especially $\sin^2\theta_{12}=1/3$ will not be spoiled and that a sizable reactor mixing angle $\theta_{13}$ and a large Dirac {\it CP}-odd phase are obtained in the mixing $U_{\rm PMNS} = V_L^{\ell\dagger}U_{\rm TBM}$. The Qin-Ma (QM) parametrization of the quark CKM matrix is a Wolfenstein-like parametrization and can be expanded in terms of the small parameter $\lambda$~\cite{Qin:2011ub}. However, unlike the original Wolfenstein parametrization \cite{Wolfenstein:1983yz}, the QM one has the advantage that its {\it CP}-odd
phase $\delta$ is manifested in the parametrization and is near maximal, {\it i.e.,} $\delta\sim 90^\circ$. This is crucial for a viable neutrino phenomenology. It should be stressed that one can also use any parametrization for the CKM matrix as a starting point. As shown in \cite{Koide}, one can adjust the phase differences in the diagonal phase matrix $P_f$ in Eq. (\ref{Vl}) in such a way that the prediction of $\sin^2\theta_{12}$ will not be considerably affected.

For $V^{\ell \dag}_{L}=V_{\rm QM}$,  the QM parametrization~\cite{Qin:2011ub,Ahn:2011ep} can be obtained
from Eq.~(\ref{Vl}) by the replacements
$s_{1}e^{i\alpha_{1}}=-(f+he^{-i\delta})\lambda^{2}~, s_{2}=f\lambda^{3}~, s_{3}=\lambda~,
\alpha_{2}=\delta~, \alpha_{3}=\delta-\pi$ :
 \begin{eqnarray}
  V^{f\dag}_{L}=P^{\ast}_{f}{\left(\begin{array}{ccc}
   1-\lambda^2/2 & \lambda e^{i\delta} & h\lambda^3 \\
   -\lambda e^{-i\delta} & 1-\lambda^2/2 & (f+h e^{-i\delta})\lambda^2 \\
   f\lambda^3 e^{-i\delta} & -(f+h e^{i\delta})\lambda^2 & 1 \\
   \end{array}\right)}+{\cal O}(\lambda^{4})~.
 \label{Vl1}
 \end{eqnarray}
On the other hand, for $V^{\ell}_{L}=V_{\rm QM}$ the QM parametrization is obtained by the replacements
$s_{1}e^{i\alpha_{1}}=(f+he^{-i\delta})\lambda^{2}~, s_{2}=h\lambda^{3}~, s_{3}=\lambda~,
\alpha_{2}=0~, \alpha_{3}=\delta$ :
 \begin{eqnarray}
  V^{f}_{L}={\left(\begin{array}{ccc}
   1-\lambda^2/2 & \lambda e^{i\delta} & h\lambda^3 \\
   -\lambda e^{-i\delta} & 1-\lambda^2/2 & (f+h e^{-i\delta})\lambda^2 \\
   f\lambda^3 e^{-i\delta} & -(f+h e^{i\delta})\lambda^2 & 1 \\
   \end{array}\right)}P_{f}+{\cal O}(\lambda^{4})~,
 \label{Vl2}
 \end{eqnarray}
where the superscript $f$ denotes $d$ (down-type quarks) or $\ell$ (charged leptons). From the global
fits to the quark mixing matrix given by~\cite{CKMfitter} we obtain
 \begin{eqnarray}
 f=0.749^{+0.034}_{-0.037}\,,\quad h=0.309^{+0.017}_{-0.012}\,,\quad \lambda=0.22545\pm 0.00065\,,
 \quad \delta=(89.6^{+2.94}_{-0.86})^\circ\,.
  \label{eq:QMfh}
 \end{eqnarray}
Because of the freedom of the phase redefinition for the quark fields, we have shown in~\cite{Ahn:2011it}
that the QM parametrization is indeed equivalent to the Wolfenstein one in the quark sector.

Finally, the leptonic mixing parameters ($\theta_{23},\theta_{12},\theta_{13},\delta_{CP}$) except Majorana phases can be expressed in terms of five parameters $\theta$ (or $\epsilon$), $\delta,f,h, \lambda$, the last four being the QM parameters in the lepton sector. If we further assume that all the QM parameters except $\delta$ have the same values in both the CKM and PMNS matrices, then only two free parameters left in the lepton mixing matrix are $\epsilon$ and $\delta$. If $\delta$ is fixed to be the same as the CKM one, then there will be only one free parameter $\epsilon$ in our calculation. In the next section, we shall study the dependence of the mixing angles $\sin^{2}\theta_{23},~\sin^{2}\theta_{12},~\sin\theta_{13}$ and
the Jarlskog invariant $J_{CP}$ on  $\delta$ and $\epsilon$.

To make our point clearer, let us summarize the reduction of the number of independent parameters in this work.
In the leptonic sector, we start with 16 free parameters (12 from the charged lepton mass matrix
$M_{\ell}^{\prime~2}$ and 4 from the neutrino mass matrix $M_{\nu}^{\prime~2}$) as shown in
Eq.~(\ref{mass2}). Among the 12 parameters from $M_{\ell}^{\prime~2}$, three phases can be rotated away
by the redefinition of the charged lepton fields.  The remaining 9 parameters correspond to three charged
lepton masses ($m_{e,\mu,\tau}$) and six angles in the charged lepton mixing matrix $V_L^{\ell}$ as shown
in Eq.~(\ref{Vl}), while the 4 parameters from $M_{\nu}^{\prime~2}$ correspond to three neutrino masses
($m_{1,2,3}$) plus one angle ($\theta$ or $\epsilon$) in the neutrino mixing matrix $U_{\nu}$ as shown in
Eq.~(\ref{Unu}) or (\ref{epsilon}). With our {\it ansatz} for $V_L^{\ell}$ discussed before, the 6 angles
in $V_L^{\ell}$ are reduced to four QM parameters ($f, h, \lambda, \delta$).
Thus, the number of parameters finally becomes five ($f, h, \lambda, \delta$ plus $\theta$ (or $\epsilon$)),
except for the six lepton masses. Under the further assumption of the QM parameters $f, h, \lambda$ having
the same values in both the CKM and PMNS matrices, these five parameters are reduced to only two ones
$\delta$ and $\epsilon$.

%%%%%%%%%%%%%%%%%%%%%%%%%%%%%%%%%%%%%%%%%%%%%%%%%%%%%%%%%%%%%%%%%%%%%%%%%%%%%%%%%%%%%%%%%%%%%%%
\section{Neutrino phenomenology}
We now proceed to discuss the low energy neutrino phenomenology with the neutrino mixing matrix $U_\nu$ (see Eq.~(\ref{Unu})) characterized by the mixing angle $\theta$ or the small parameter $\epsilon$ and the charged lepton mixing matrix $U_{L}=V_{C}V^{\ell}_{L}$ in which $V^{\ell}_{L}$ is assumed to have the similar expression as the QM parametrization~\cite{Qin:2011ub,Ahn:2011ep} given by $V^{\dag}_{\rm QM}$ or $V_{\rm QM}$ (see Eq.~(\ref{Vl1}) and Eq.~(\ref{Vl2}), respectively).  The lepton mixing matrix thus has the form
\begin{eqnarray}
U_{\rm PMNS} =\left\{
 \begin{array}{ll}
  V_{\rm QM}U_{\rm TBM}W & {\rm for}~ V_L^{\ell\dagger}=V_{\rm QM}, \hbox{} \\
  V_{\rm QM}^{\dag}U_{\rm TBM}W & {\rm for}~ V_L^{\ell}=V_{\rm QM}. \hbox{}
  \end{array} \right.
\end{eqnarray}
Therefore, the corrections to the TBM matrix within our framework arise from the charged lepton mixing
matrix $V_L^\ell$ characterized by the parameters $f,h,\lambda,\delta$ and the matrix $W$ specified by
the parameter $\epsilon$ whose size is strongly constrained by the recent T2K data. Indeed, the
parameters $\lambda,~f,~h$ and $\delta$ in the lepton sector are {\it a priori} not necessarily the
same as that in the quark sector. Hereafter, we shall use the central values in Eq.~(\ref{eq:QMfh}) of
the parameters $(\lambda,~f,~h)$ for our numerical calculations.

In the following we consider both cases:
%%%%%%%%%%%%%%%%%%%%%%%%%%%%%%%%%%%%%%%%%%
\vskip 0.4cm
{\bf (i)} $V^{\ell\dag}_{L}=V_{\rm QM}$
\vskip 0.5cm

\begin{figure}[t]
%\vspace*{-5.0cm}
\hspace*{-2cm}
\begin{minipage}[t]{6.0cm}
\epsfig{figure=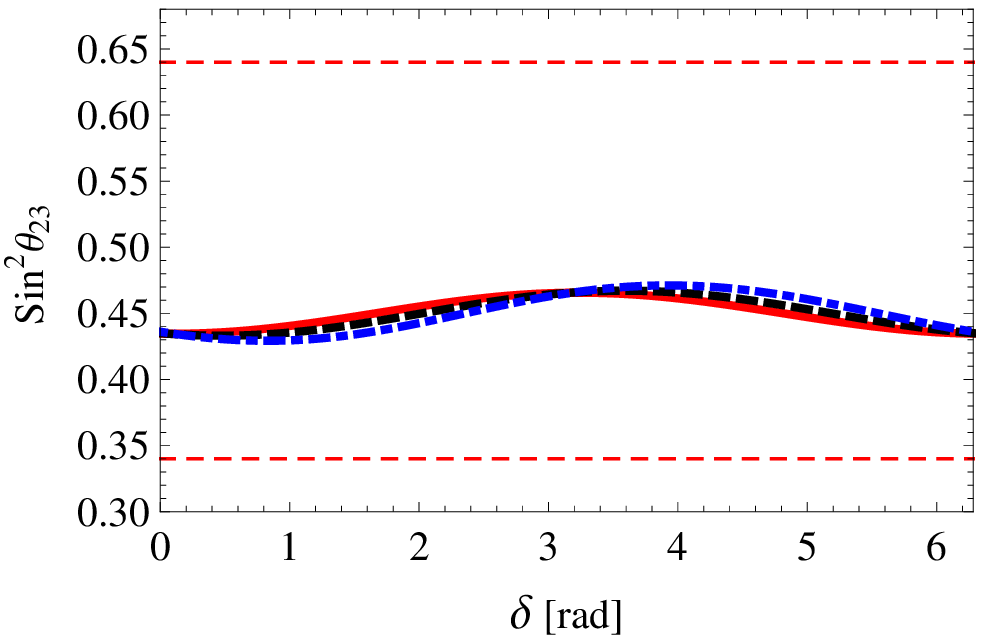,width=6.5cm,angle=0}
\end{minipage}
\hspace*{2.0cm}
\begin{minipage}[t]{6.0cm}
\epsfig{figure=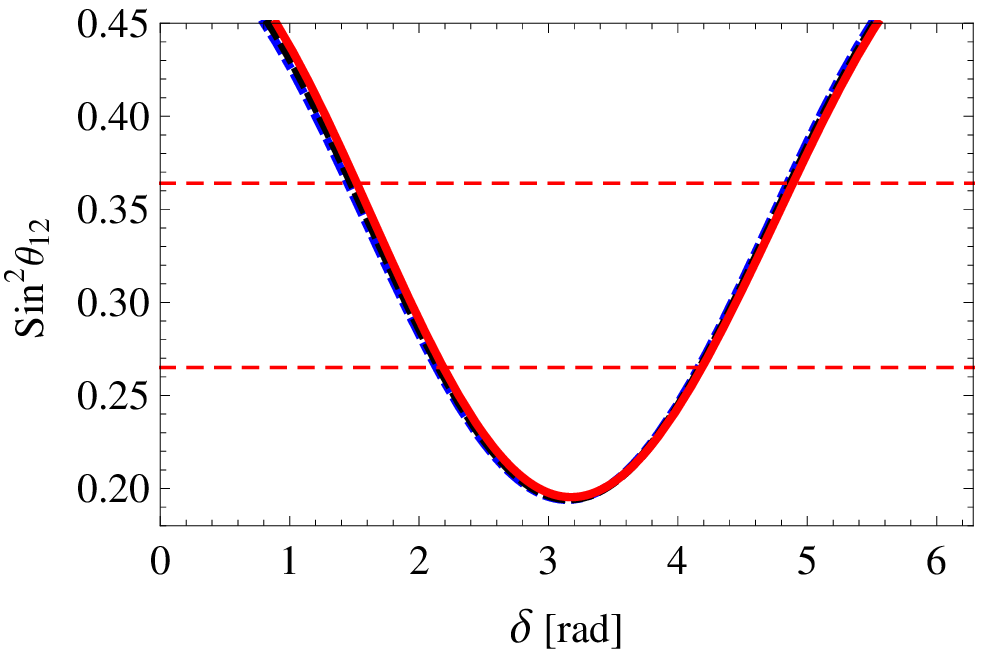,width=6.5cm,angle=0}
\end{minipage}
\vspace*{-1.0cm} \hspace*{-2cm}
\begin{minipage}[t]{6.0cm}
\epsfig{figure=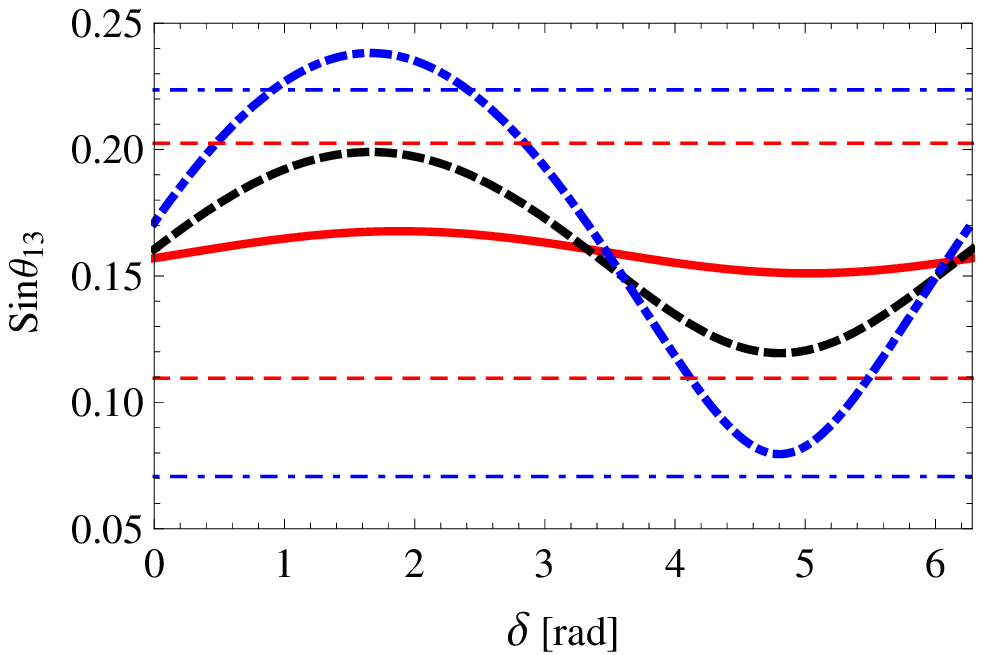,width=6.5cm,angle=0}
\end{minipage}
\hspace*{2.0cm}
\begin{minipage}[t]{6.0cm}
\epsfig{figure=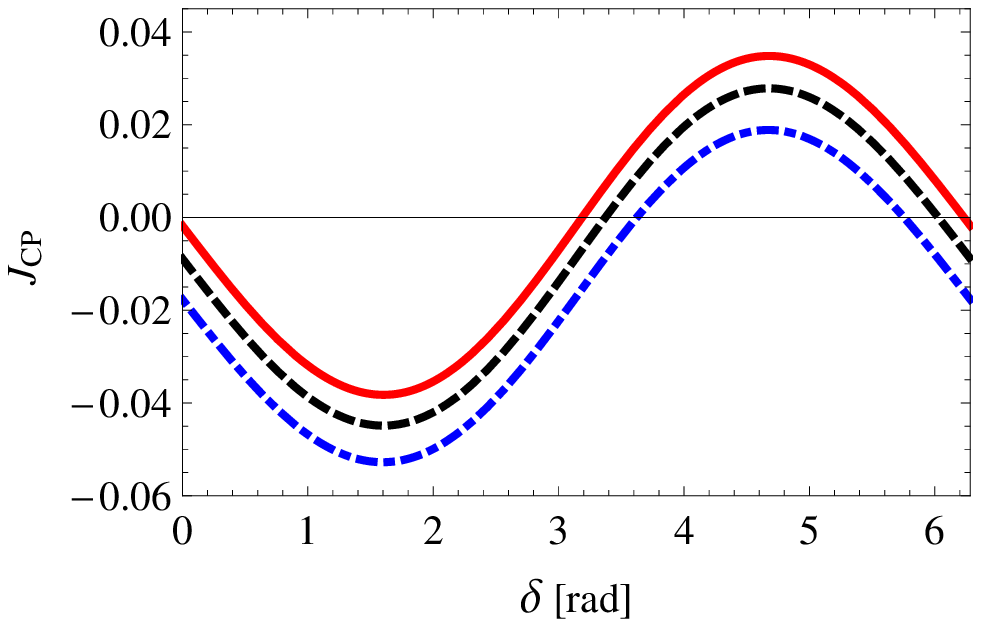,width=6.5cm,angle=0}
\end{minipage}
\vspace*{1.0cm}
\caption{\label{Fig1} Mixing angles $\sin^{2}\theta_{23},~\sin^{2}\theta_{12},~\sin\theta_{13}$ and
the Jarlskog invariant $J_{CP}$ as a function of the {\it CP}-odd phase $\delta$ for case (i), where
the solid, dashed and dot-dashed curves are for $\epsilon=0.01$, 0.05 and 0.1, respectively, the
horizontal dashed lines denote the upper and lower bounds in $3\sigma$ set by Fogli {\it et al.}
(see Eq.~(\ref{exp})) and use of $\lambda=0.2254,~f=0.749,~h=0.309$ has been made. In the plot of
$\sin\theta_{13}$ versus $\delta$, the horizontal dot-dashed lines represent the present data of T2K
Collaboration for the normal neutrino mass hierarchy, see Eq.~(\ref{T2K}).}
\end{figure}

%\subsection{For $V^{\ell\dag}_{L}=V_{\rm CKM}$}
With the help of Eqs.~(\ref{HPS}) and (\ref{Vl1}), the leptonic mixing matrix corrected by the
replacements $V_{C}\rightarrow U_{L}=V_CV^{\ell}_L=V_{C}V^{\dag}_{\rm QM}$ and
$U_{\nu}(\pi/4)\rightarrow U_{\nu}(\pi/4+\epsilon)$, can be written, up to order of $\lambda^{3}$ and
$\epsilon^{2}$, as
 \begin{eqnarray}
 U_{\rm PMNS}^{\rm (i)}&=&U^{\dag}_{L}~U_{\nu}(\pi/4+\epsilon)= V_{\rm QM} U_{\rm TBM}W
 \nonumber\\
 &=&U_{\rm TBM}+\epsilon{\left(\begin{array}{ccc}
 -\frac{\epsilon}{2}\sqrt{\frac{2}{3}} &  0 & \sqrt{\frac{2}{3}} \\
 \frac{i}{\sqrt{2}}+\frac{\epsilon}{2\sqrt{6}} &  0 &  -\frac{1}{\sqrt{6}}+\frac{i\epsilon}{2\sqrt{2}} \\
 -\frac{i}{\sqrt{2}}+\frac{\epsilon}{2\sqrt{6}} &  0 & -\frac{1}{\sqrt{6}}-\frac{i\epsilon^{2}}{2\sqrt{2}}
 \end{array}\right)} \nonumber\\
 &+&\lambda{\left(\begin{array}{ccc}
 -\frac{e^{i\delta}+\lambda+h\lambda^{2}}{\sqrt{6}} & \frac{e^{i\delta}-\frac{\lambda}{2}-h\lambda^{2}}{\sqrt{3}} & -\frac{i(e^{i\delta}-h\lambda^{2})}{\sqrt{2}} \\
 \frac{-2 e^{-i\delta}+(\frac{1}{2}-f-he^{-i\delta})\lambda}{\sqrt{6}} & -\frac{e^{-i\delta}+(\frac{1}{2}-f-he^{-i\delta})\lambda}{\sqrt{3}} &  \frac{i(\frac{1}{2}+f+he^{-i\delta})\lambda}{\sqrt{2}} \\
 \frac{(f+he^{i\delta})\lambda+2fe^{-i\delta}\lambda^{2}}{\sqrt{6}} &  -\frac{(f+he^{i\delta})\lambda+fe^{-i\delta}\lambda^{2}}{\sqrt{3}} &  \frac{i(f+he^{i\delta})\lambda}{\sqrt{2}}
 \end{array}\right)}  \\
 &-&\lambda\epsilon{\left(\begin{array}{ccc}
 -\frac{i(e^{i\delta}-h \lambda^{2})}{\sqrt{2}}-\epsilon\frac{e^{i\delta}+\lambda+h\lambda^{2}}{2\sqrt{6}} & 0 & \frac{e^{i\delta}+\lambda+h\lambda^{2}}{\sqrt{6}}-\epsilon
 \frac{i(e^{i\delta}-h\lambda^{2})}{2\sqrt{2}} \\
 \frac{i(\frac{1}{2}+f+he^{-i\delta})\lambda}{\sqrt{2}}-\epsilon
 \frac{2e^{-i\delta}+(f+he^{-i\delta}-\frac{1}{2})\lambda}{2\sqrt{6}} & 0 & \frac{2e^{-i\delta}+(f+he^{-i\delta}-\frac{1}{2})\lambda}{\sqrt{6}}
 +\epsilon\frac{i(f+he^{-i\delta}+\frac{1}{2})\lambda}{2\sqrt{2}} \\
 \frac{i(f+he^{i\delta})\lambda}{\sqrt{2}}+\epsilon\frac{(f+he^{i\delta})
 \lambda+2fe^{-i\delta}\lambda^{2}}{2\sqrt{6}} & 0 & -\frac{(f+he^{i\delta})\lambda+2\lambda^{2}fe^{i\delta}}{\sqrt{6}}
 +\epsilon\frac{i(f+he^{i\delta})\lambda}{2\sqrt{2}}
 \end{array}\right)}~. \nonumber
 \label{leptonA}
 \end{eqnarray}
%which indicates the TBM at zeroth order and the observed deviation from TBM as small expansion parameters $\epsilon$ and $\lambda$.
Note that $U_{\rm PMNS}^{\rm (i)}$ here contains five independent parameters ($\lambda,h,f,\delta$ and
$\epsilon$).\footnote{Our previous work \cite{Ahn:2011ep} corresponds to case (i) with $\epsilon=0$\ .}
By rephasing the lepton and neutrino fields $e \to e \,e^{i\alpha_{1}}$,
$\mu \to \mu \,e^{i\beta_{1}}$, $\tau \to \tau \,e^{i\beta_{2}}$ and
$\nu_{2} \to \nu_{2} \,e^{i(\alpha_{1}-\alpha_{2})}$, the PMNS matrix is recast to
 \begin{eqnarray}
  U_{\rm PMNS}=
 {\left(\begin{array}{ccc}
 |U_{e1}| & |U_{e2}| & |U_{e3}|e^{-i(\alpha_{1}-\alpha_{3})} \\
 U_{\mu1}e^{-i\beta_{1}} & U_{\mu2}e^{i(\alpha_{1}-\alpha_{2}-\beta_{1})} &  |U_{\mu3}| \\
 U_{\tau1}e^{-i\beta_{2}} & U_{\tau2}e^{i(\alpha_{1}-\alpha_{2}-\beta_{2})} & |U_{\tau3}|
 \end{array}\right) P_{\nu}}~,
 \label{PMNS2}
 \end{eqnarray}
where $U_{\alpha j}$ is an element of the PMNS matrix  with $\alpha=e,\mu,\tau$ corresponding to the
lepton flavors and $j=1,2,3$ to the light neutrino mass eigenstates.
In Eq.~(\ref{PMNS2}) the phases defined as $\alpha_{1} = \arg(U_{e1})$, $\alpha_{2} = \arg(U_{e2})$,
$\alpha_{3} = \arg(U_{e3})$, $\beta_{1} = \arg(U_{\mu3})$ and $\beta_{2} = \arg(U_{\tau3})$ have the
expressions:
 \begin{eqnarray}
  \alpha_{1}&=&\tan^{-1}\left(\frac{\lambda\{\sqrt{3}(\epsilon^{2}-2)\sin\delta+6\epsilon\cos\delta
  -6h\epsilon\lambda^{2}\}}{\sqrt{3}(2-\epsilon^{2})(2-\lambda^{2}-h\lambda^{3})
  +\sqrt{3}(\epsilon^{2}-2)\lambda\cos\delta-6\epsilon\lambda\sin\delta}\right)~\ , \nonumber\\
  \alpha_{2}&=&\tan^{-1}\left(\frac{\lambda\sin\delta}{1+\lambda\cos\delta
  -\frac{\lambda^{2}}{2}+h\lambda^{3}}\right)~ \ , \nonumber\\
  \alpha_{3}&=&\tan^{-1}\left(\frac{\lambda\{2\sqrt{3}\epsilon\sin\delta+3(2-\epsilon^{2})\cos\delta
  -3h(2-\epsilon^{2})\lambda^{3}\}}{3\lambda(\epsilon^{2} -2)\sin\delta-2\sqrt{3}\epsilon(2-\lambda^{2}-\lambda\cos\delta-h\lambda^{3})}\right)~\ , \nonumber\\
  \beta_{1}
  &=&\tan^{-1}\left(\frac{3(2-\epsilon^{2})(2-\lambda^{2}-2f\lambda^{2})-6h(2-\epsilon^{2})
  \lambda^{2}\cos\delta-4\sqrt{3}\epsilon\lambda(2+h\lambda)\sin\delta}{2\sqrt{3}\epsilon
  (2-\lambda^{2}+2f\lambda^{2})+4\sqrt{3}\epsilon\lambda(2+h\lambda)\cos\delta-6h(2-\epsilon^{2})
  \lambda^{2}\sin\delta}\right)\ , \nonumber\\
  \beta_{2}&=&\tan^{-1}
  \left(\frac{3(2-\epsilon^{2})(1+f\lambda^{2})+3h\lambda^{2}(2-\epsilon^{2})\cos\delta+2\sqrt{3}
  \epsilon\lambda^{2}(h-2f\lambda)\sin\delta}{2\sqrt{2}\epsilon(f\lambda^{2}-1)+2\sqrt{3}\epsilon
  \lambda^{2}(h+2f\lambda)\cos\delta-3h\lambda^{2}(2-\epsilon^{2})\sin\delta}\right)~.
  \label{mixing elements2}
 \end{eqnarray}
>From Eq.~(\ref{PMNS2}), the neutrino mixing parameters can be displayed as
  \begin{eqnarray}
  \sin^{2}\theta_{12}&=&\frac{|U_{e2}|^{2}}{1-|U_{e3}|^{2}}~,\qquad\qquad\quad
   \sin^{2}\theta_{23}=\frac{|U_{\mu3}|^{2}}{1-|U_{e3}|^{2}}~,\nonumber\\
  \sin\theta_{13}&=&|U_{e3}|~,\qquad\qquad\qquad\quad ~\delta_{CP}=\alpha_{1}-\alpha_{3}~.
 \label{mixing1}
 \end{eqnarray}
It follows from Eqs.~(\ref{leptonA}) and (\ref{mixing1}) that the solar neutrino mixing angle
$\theta_{12}$ can be approximated, up to order $\lambda^3$ and $\epsilon^{2}$, as
 \begin{eqnarray}
  \sin^{2}\theta_{12}&\simeq&\frac{1}{3}+\frac{2\epsilon^{2}}{9}+\frac{2\lambda}{3}
  \left(\cos\delta+\frac{\epsilon\sin\delta}{\sqrt{3}}+\frac{\epsilon^2\cos\delta}{3}\right)\nonumber\\
  &+&\frac{\lambda^{2}}{3}\left(\frac{1}{2}+\frac{2\epsilon\sin2\delta}{\sqrt{3}}
  -\frac{\epsilon^{2}}{3}(3+4\cos^{2}\delta)\right)~\nonumber\\
  &+&\frac{\lambda^{3}}{3}\left(2h-\frac{\epsilon\sin\delta}{\sqrt{3}}
  +\frac{\epsilon^{2}}{3}(2h-7\cos\delta)\right) \ .
 \label{Sol}
 \end{eqnarray}
This indicates that the deviation from $\sin^{2}\theta_{12}=1/3$ becomes small when $\cos\delta$
approaches to zero and the magnitude of $\epsilon$ is less than $\lambda$. Since it is the first column
of $V_L^\ell$ that makes the major contribution to $\sin^{2}\theta_{12}$, this explains why we need a
phase of order $90^\circ$ for the element  $(V_L^\ell)_{21}$: When $|\sin\delta|\approx1$, the present
data of the solar mixing angle can be accommodated even for a large $|\epsilon|$ (but less than $\lambda$).
The behavior of $\sin^{2}\theta_{12}$ as a function of $\delta$ is plotted in Fig.~\ref{Fig1} where the
horizontal dashed lines denote the upper and lower bounds of the experimental data in $3\sigma$ ranges.
The allowed regions for $\delta$ (in radian) lie in the ranges of $1.45\lesssim\delta\lesssim2.17$ and
$4.17\lesssim\delta\lesssim4.91$\,, recalling that the QM phase is $\delta_{\rm QM}=1.56$\,.
%In Fig.~\ref{Fig1} the plot $\sin^{2}\theta_{12}$ versus $\delta$ shows that the present data is consistent with $|\sin\delta|\approx1$, even in the case of $\epsilon=0.01$ (solid curve), $\epsilon=0.05$ (dashed curve) and $\epsilon=0.1$ (dot-dashed curve).

Likewise, the atmospheric neutrino mixing angle $\theta_{23}$ comes out as
 \begin{eqnarray}
   \sin^{2}\theta_{23}&\simeq&\frac{1}{2}-\frac{\epsilon\lambda}{\sqrt{3}}\left(\sin\delta
   -\frac{\epsilon\cos\delta}{\sqrt{3}}\right)\nonumber\\
   &-&\lambda^{2}\left(\frac{1}{4}+f+h\cos\delta+\epsilon\frac{2h\sin\delta}{\sqrt{3}}
   +\frac{2\epsilon^{2}}{3}(1-f-h\cos\delta-\cos2\delta)\right)~\nonumber\\
   &-&\lambda^{3}\epsilon\left(\frac{\sin\delta}{2\sqrt{3}}(3+4h\cos\delta)-\epsilon
   \left[\frac{3-8f}{6}\cos\delta+h-2h\cos^{2}\delta\right]\right) \ .
 \label{Atm}
 \end{eqnarray}
Fig. \ref{Fig1} shows a small deviation from the TBM atmospheric mixing angle with $\theta_{23}<45^\circ$
for $0<|\epsilon|<\lambda$.
%It is thus crucial to have precise measurements of the atmospheric mixing angle in future experiments to see whether $\theta_{23}\leq45^{\circ}$ or $\theta_{23}\geq45^{\circ}$ in order to test different scenarios.
Owing to the absence of corrections to the first order of $\lambda$ or $\epsilon$ in Eq.~(\ref{Atm}),
the deviation from the maximal mixing of $\theta_{23}$ comes mainly from the terms associated with
$\lambda^{2}$ or $\epsilon\lambda$.
Especially, for $\sin\delta\approx1$ we have the approximation
$\sin^{2}\theta_{23}-\frac{1}{2}\approx-\frac{\epsilon\lambda}{\sqrt{3}}-\lambda^{2}(f+\frac{1}{4})$,
which implies $\sin^2\theta_{23}<1/2$ for $0<|\epsilon|<\lambda$. We see from Fig.~\ref{Fig1} that
$\sin^2\theta_{23}$ lies in the ranges $0.43<\sin^{2}\theta_{23}<0.45$ for $0\leq|\epsilon|\lesssim0.1$\,.

The reactor mixing angle $\theta_{13}$ now reads
 \begin{eqnarray}
  \sin^2\theta_{13}&=&\frac{2\epsilon\lambda\sin\delta}{\sqrt{3}}+\frac{2\epsilon^{2}}{3}
  (1-\lambda\cos\delta) \nonumber \\ &+&\lambda^{2}\left(\frac{1}{2}-\epsilon^{2}\right)-\lambda^{3}\epsilon
  \left(\frac{\sin\delta}{\sqrt{3}}+\frac{2h\epsilon}{3}-\frac{\epsilon\cos\delta}{3}\right)~.
 \label{Reactor}
 \end{eqnarray}
Evidently, $\sin\theta_{13}$ depends considerably on the parameters $\lambda$ and $\epsilon$. Thus, we
have a non-vanishing $\theta_{13}$  with a central value of $\sin\theta_{13}=\lambda/\sqrt{2}$ or
$\theta_{13}=9.2^\circ$ for $\epsilon=0$~\cite{Ahn:2011ep}.
Note that the size of the unknown parameter $\epsilon$ is constrained by the plot of $\sin\theta_{13}$
versus $\delta$ in Fig.~\ref{Fig1} where the horizontal dot-dashed lines represent the present T2K data
for the normal neutrino mass hierarchy. For a negative value of $\epsilon$, the plot for $\sin\theta_{13}$
versus $\delta$ is flipped upside-down. Assuming $\rho>0$, we see from Eq.~(\ref{masssquare}) that a
positive (negative) value of $\epsilon$ leads to a normal (inverted) neutrino mass spectrum. For example,
we find $\frac{\lambda}{\sqrt{2}}\leq\sin\theta_{13}\lesssim0.22$
($0.07\lesssim\sin\theta_{13}\leq\frac{\lambda}{\sqrt{2}}$) for $\delta=1.56$ and $\epsilon\leq0.08$
($\epsilon\geq-0.11$)\ .

Leptonic {\it CP} violation can be detected through the neutrino oscillations which are sensitive to
the Dirac {\it CP}-phase $\delta_{CP}$, but insensitive to the Majorana phases in
$U_{\rm PMNS}$~\cite{Branco:2002xf}.
It follows from Eqs.~(\ref{mixing elements2}) and (\ref{mixing1}) that the Dirac phase
$\delta_{CP}=\alpha_1-\alpha_3$ has the expression
 \begin{eqnarray}
  \delta_{CP}= \tan^{-1}\left( \frac{\sqrt{3}\lambda\{-2\cos\delta+\lambda\cos2\delta+\lambda^2
  \cos\delta+\sqrt{3}\epsilon\lambda\sin2\delta\}}{2\sqrt{3}\lambda\sin
  \delta\{1-\lambda\cos\delta-\frac{\lambda^2}{2}\}-4\epsilon\{1-\lambda
  \cos\delta-\frac{3}{2}\lambda^2\cos^2\delta-\lambda^3(h-\frac{\cos\delta}{2})\}} \right) ~, \nonumber \\
 \label{DiracCP1}
 \end{eqnarray}
where terms of order $\epsilon^3,\lambda^4,\epsilon^2\lambda^2$ have been neglected in both numerator and denominator.
Assuming $\rho>0$, we show in Table~\ref{DiracCP11} the predictions for $\delta_{P}$ and $\theta_{13}$ as a function of $\epsilon$ , where we have used the central values of Eq.~(\ref{eq:QMfh}).
\begin{center}
\begin{table}[h]
\caption{\label{DiracCP11} Predictions of $\delta_{CP}$ and $\theta_{13}$ as a function of $\epsilon$ in the case of $V^{\ell\dag}_{L}=V_{\rm QM}$.}
\begin{ruledtabular}
\begin{tabular}{ccc}
 $\epsilon$ & $\delta_{CP}~[{\rm deg.}]$ & $\theta_{13}~[{\rm deg.}]$ \\
\hline
 $-0.012\sim0.08$ & $-173.6\sim-169$ & $9.4\sim5.8$ \\
 $-0.11\sim-0.012$ & $184.6\sim186.4$ & $14.4\sim9.4$
\end{tabular}
\end{ruledtabular}
\end{table}
\end{center}
To see how the parameters are correlated with low energy {\it CP} violation measurable through
neutrino oscillations, let us consider the leptonic {\it CP} violation parameter defined through the
Jarlskog invariant $J_{CP}\equiv{\rm Im}[U_{e1}U_{\mu2}U^{\ast}_{e2}U^{\ast}_{\mu1}]
 =\frac{1}{8}\sin2\theta_{12}\sin2\theta_{23} \sin2\theta_{13}\cos\theta_{13}
 \sin\delta_{CP}$~\cite{Jarlskog:1985ht} which is expressed as
 \begin{eqnarray} \label{JCP1}
  J_{CP}&=&-\frac{\epsilon}{3\sqrt{3}}-\frac{\lambda}{6}\left(\sin\delta+\epsilon\frac{4\cos\delta}
  {\sqrt{3}}-2\epsilon^{2}\sin\delta\right) \\
  &-&\frac{\lambda^{2}}{9}\left(\sin\delta(h+\cos\delta)-\epsilon\sqrt{3}(1+f-\cos2\delta+h\cos\delta)
  -\epsilon^{2}(h+4\cos\delta)\sin\delta\right). \nonumber
 \end{eqnarray}
We see from the above equation that $J_{CP}$ is strongly correlated with $\epsilon$ and $\delta$ for the
fixed values of $\lambda,~h$ and $f$. As long as $\epsilon\neq0$ (associated with the neutrino part) or
$\lambda\neq0$ (associated with the charged lepton part), $J_{CP}$ has a non-vanishing value, indicating
a signal of {\it CP} violation. Eq.~(\ref{JCP1}) could be approximated as
$J_{CP}\approx-\frac{\epsilon}{3\sqrt{3}}-\frac{\lambda}{6}\sin\delta$.
The behavior of $J_{CP}$ is plotted in Fig.~\ref{Fig1} as a function of $\delta$. When $\sin\delta\approx1$,
it is reduced to $J_{CP}\approx-\frac{\epsilon}{3\sqrt{3}}-\frac{\lambda}{6}
\leq-\frac{\lambda}{6}~(\geq-\frac{\lambda}{6})$ for $\epsilon>0~(\epsilon<0)$. Assuming $\rho>0$, we
find $-0.050\lesssim J_{CP}\lesssim-0.037$ ($-0.037\lesssim J_{CP}\lesssim-0.017$) for $\epsilon\leq0.08$
($\epsilon\geq-0.11$) and $\delta=1.56$\,.

%%%%%%%%%%%%%%%
%    Fig 1    %
%%%%%%%%%%%%%%%

%%%%%%%%%%%%%%%%%%%%%%%%%%%%%%%%%%%%%%%%%%
\vskip 0.4cm
{\bf (ii)} $V^{\ell}_{L}=V_{\rm QM}$
\vskip 0.5cm

\begin{figure}[t]
%\vspace*{-5.0cm}
\hspace*{-2cm}
\begin{minipage}[t]{6.0cm}
\epsfig{figure=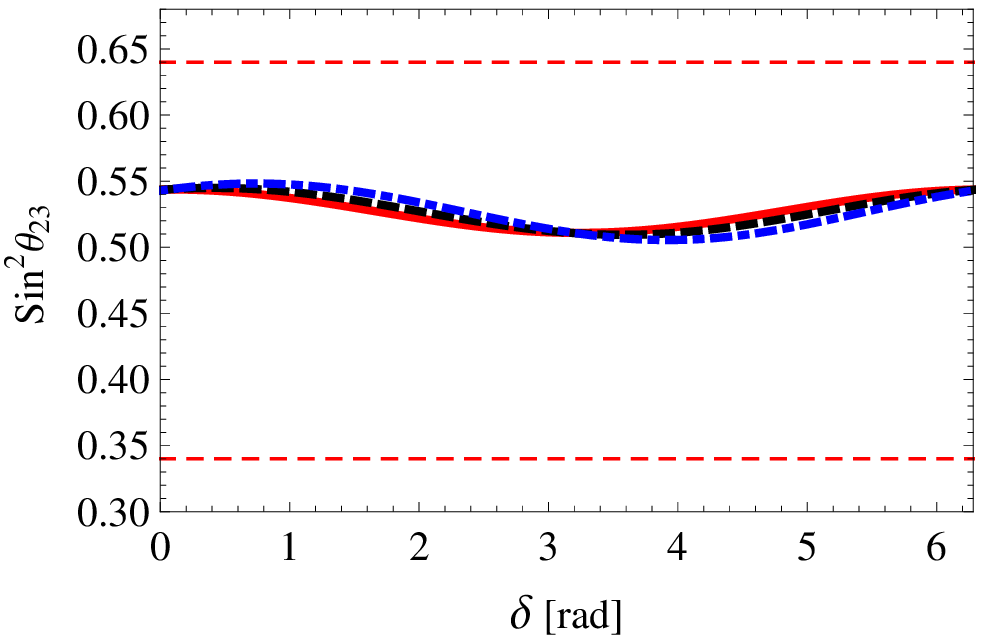,width=6.5cm,angle=0}
\end{minipage}
\hspace*{2.0cm}
\begin{minipage}[t]{6.0cm}
\epsfig{figure=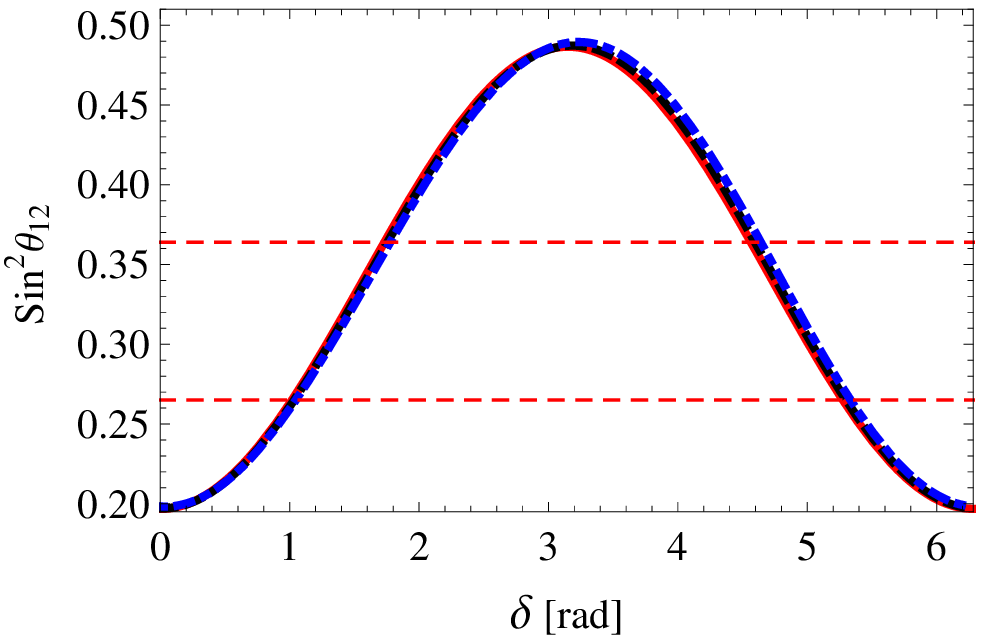,width=6.5cm,angle=0}
\end{minipage}
\vspace*{-1.0cm} \hspace*{-2cm}
\begin{minipage}[t]{6.0cm}
\epsfig{figure=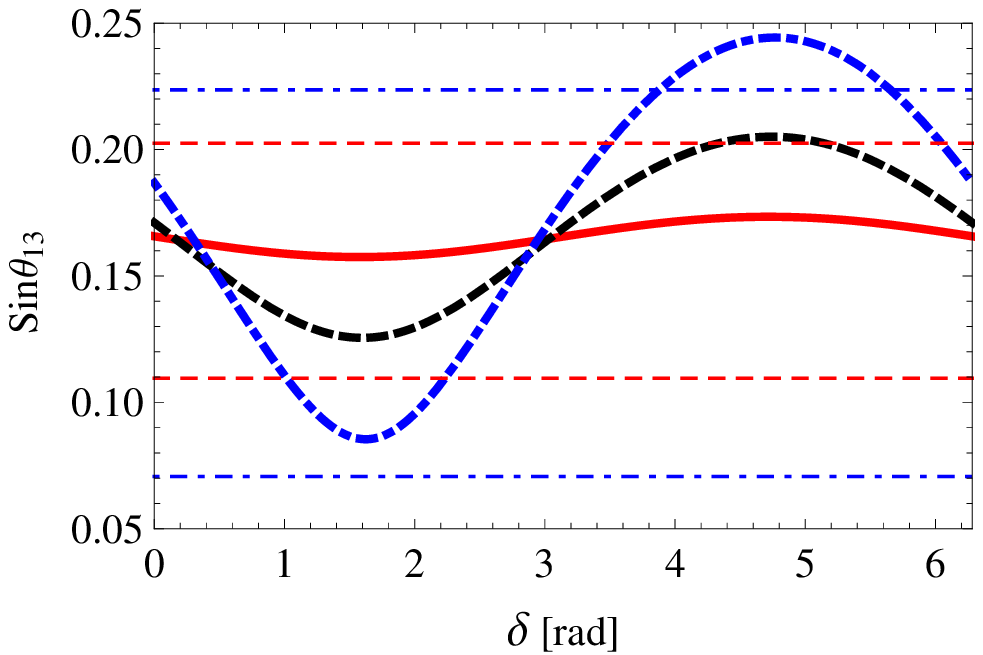,width=6.5cm,angle=0}
\end{minipage}
\hspace*{2.0cm}
\begin{minipage}[t]{6.0cm}
\epsfig{figure=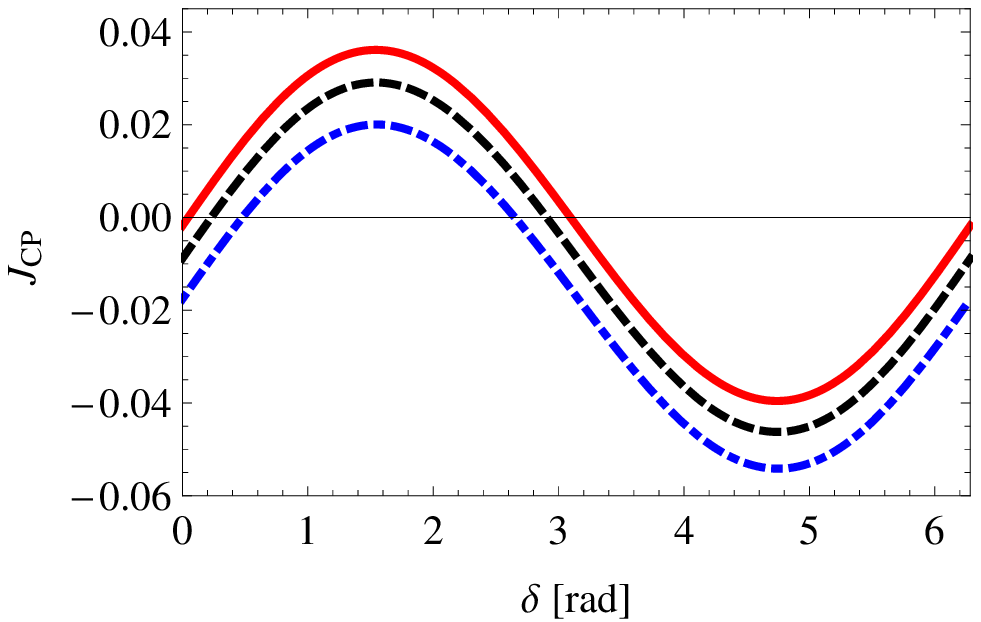,width=6.5cm,angle=0}
\end{minipage}
\vspace*{1.0cm}
\caption{\label{Fig2} Same as Fig. \ref{Fig1} except for case (ii).}
\end{figure}

The resulting leptonic mixing matrix in this case can be expressed, up to order of $\lambda^{3}$ and
$\epsilon^{2}$, as
 \begin{eqnarray}  \label{leptonB}
 U_{\rm PMNS}^{\rm (ii)}&=&U^{\dag}_{L}~U_{\nu}(\pi/4+\epsilon)= V^{\dag}_{\rm QM} U_{\rm TBM}W \nonumber\\
 &=&U_{\rm TBM}+{\left(\begin{array}{ccc}
 -\frac{\epsilon^{2}}{2}\sqrt{\frac{2}{3}} &  0 & \epsilon\sqrt{\frac{2}{3}} \\
 \frac{i\epsilon}{\sqrt{2}}+\frac{\epsilon^{2}}{2\sqrt{6}} &  0 &  -\frac{\epsilon}{\sqrt{6}}+\frac{i\epsilon^{2}}{2\sqrt{2}} \\
 -\frac{i\epsilon}{\sqrt{2}}+\frac{\epsilon^{2}}{2\sqrt{6}} &  0 &  -\frac{\epsilon}{\sqrt{6}}-\frac{i\epsilon^{2}}{2\sqrt{2}}
 \end{array}\right)}\nonumber\\
 &+&\lambda{\left(\begin{array}{ccc}
 \frac{e^{i\delta}-\lambda-fe^{i\delta}\lambda^{2}}{\sqrt{6}} & -\frac{e^{i\delta}+\frac{1}{2}\lambda-fe^{i\delta}\lambda^{2}}{\sqrt{3}} & \frac{ie^{i\delta}(1+f\lambda^{2})}{\sqrt{2}} \\
 \frac{2 e^{-i\delta}+(\frac{1}{2}+f+he^{-i\delta})\lambda}{\sqrt{6}} & \frac{e^{-i\delta}-(\frac{1}{2}+f+he^{-i\delta})\lambda}{\sqrt{3}} &  \frac{i(\frac{1}{2}-f-he^{-i\delta})\lambda}{\sqrt{2}} \\
 -\frac{(f+he^{i\delta})\lambda-2\lambda^{2}h}{\sqrt{6}} &  \frac{(f+he^{i\delta})\lambda+h\lambda^{2}}{\sqrt{3}} &  -\frac{i(f+he^{i\delta})\lambda}{\sqrt{2}}
 \end{array}\right)}  \\
 &+&
 \lambda\epsilon{\left(\begin{array}{ccc}
 -\frac{ie^{i\delta}(1+f\lambda^{2})}{\sqrt{2}}+\frac{-e^{i\delta}\epsilon
 +\lambda\epsilon+fe^{i\delta}\lambda^{2}\epsilon}{2\sqrt{6}} & 0 & \frac{e^{i\delta}-\lambda-fe^{i\delta}\lambda^{2}}{\sqrt{6}}
 +\frac{-e^{i\delta}\epsilon-ife^{i\delta}\lambda^{2}\epsilon}{2\sqrt{2}} \\
 \frac{i(f+he^{-i\delta}-\frac{1}{2})\lambda}{\sqrt{2}}-\frac{2e^{-i\delta}
 \epsilon+(\frac{1}{2}+f+he^{-i\delta})\lambda\epsilon}{2\sqrt{6}} & 0 & \frac{2e^{-i\delta}+(\frac{1}{2}+f+he^{-i\delta})\lambda}{\sqrt{6}}
 +\frac{i(f+he^{-i\delta}-\frac{1}{2})\lambda\epsilon}{2\sqrt{2}} \\
 \frac{i(f+he^{i\delta})\lambda}{\sqrt{2}}+\frac{(f+he^{i\delta})
 \lambda\epsilon-2h\lambda^{2}\epsilon}{2\sqrt{6}} & 0 &  -\frac{(f+he^{i\delta})\lambda+2\lambda^{2}h}{\sqrt{6}}+\frac{i(f+he^{i\delta})\lambda\epsilon}{2\sqrt{2}}
 \end{array}\right)}~. \nonumber
 \end{eqnarray}
Just as in case ({i}), the exact TBM is recovered when both $\epsilon$ and $\lambda$ go to zero.
With the help of Eqs.~(\ref{mixing1}) and (\ref{leptonB}), the solar neutrino mixing angle $\theta_{12}$
can be approximated as
 \begin{eqnarray}
  \sin^{2}\theta_{12}&\simeq&\frac{1}{3}+\frac{2\epsilon^{2}}{9}-\frac{2\lambda}{3}
  \left(\cos\delta+\frac{\epsilon\sin\delta}{\sqrt{3}}+\frac{\epsilon^2\cos\delta}{3}\right)\nonumber\\
  &+&\frac{\lambda^{2}}{3}\left(\frac{1}{2}+\frac{2\epsilon\sin2\delta}{\sqrt{3}}
  -\frac{\epsilon^{2}}{3}(3+4\cos^{2}\delta)\right)~\nonumber\\
  &+&\frac{\lambda^{3}}{3}\left(2f\cos\delta+\frac{\epsilon\sin\delta}{\sqrt{3}}(1-2f)
  +\frac{\epsilon^{2}\cos\delta}{3}(2f+7)\right)\ ,
 \label{Sol2}
 \end{eqnarray}
which leads to, as in case ({i}), a tiny deviation from $\sin^{2}\theta_{12}=1/3$ when $\cos\delta \to 0$
and $\lambda>|\epsilon|$. As expected, since the second column related to $\epsilon$ in the matrix
Eq.~(\ref{leptonB}) is zero, the solar mixing angle is not affected to the first order of $\epsilon$.
Because of a minus sign in front of the $\lambda\cos\delta$ term, which constitutes the major correction
to $\sin\theta_{12}$,  the plot of $\sin^{2}\theta_{12}$ versus $\delta$ (see Fig.~\ref{Fig2}) is turned
upside-down, contrary to case (i).
When $\sin\delta \approx 1$, the present data of the solar mixing angle are well accommodated even for
a large $|\epsilon|$ (but less than $\lambda$).
%In Fig.~\ref{Fig2} the plot $\sin^{2}\theta_{12}$ versus $\delta$ shows that the present data is
%consistent with $|\sin\delta|\approx1$, even in the case of $\epsilon=0.01$ (solid curve), $\epsilon=0.05$
%(dashed curve) and $\epsilon=0.1$ (dot-dashed curve). Here the horizontal dashed lines denote the upper
%and lower bounds at $3\sigma$ in Eq.~(\ref{exp}),
The allowed regions for $\delta$ lie in the ranges of  $1.0<\delta<1.7$ and $4.5<\delta<5.3$\,.
This indicates that when the {\it CP}-odd phase $\delta$ is near maximal, the data of $\sin^2\theta_{12}$
can be easily accommodated in case (ii) but only marginally in case (i). Hence, the precise measurements
of the solar mixing angle in future experiments will tell which scenario is more preferable.

%%%%%%%%%%%%%%%
%    Fig 2    %
%%%%%%%%%%%%%%%

>From Eqs.~(\ref{mixing1}) and (\ref{leptonB}), the atmospheric neutrino mixing angle $\theta_{23}$ comes
out as
 \begin{eqnarray}
   \sin^{2}\theta_{23}&\simeq&\frac{1}{2}+\frac{\epsilon\lambda}{\sqrt{3}}
   \left(\sin\delta+\frac{\epsilon\cos\delta}{\sqrt{3}}(1-2\sqrt{3})\right)\nonumber\\
   &-&\lambda^{2}\left(\frac{1}{4}-f-h\cos\delta-\epsilon\frac{2h\sin\delta}{\sqrt{3}}
   +\frac{\epsilon^{2}}{3}(2\sin^{2}\delta+f+h\cos\delta)\right)~\nonumber\\
   &+&\frac{\lambda^{3}\epsilon}{2}\Bigg(\frac{\sqrt{3}}{3}\sin\delta(3-2f-4h\cos\delta)
   \nonumber \\
   &-& \frac{\epsilon}{3}\left[(1+2\sqrt{3}-6f-4h\cos\delta)\cos\delta+8h\sin^{2}\delta-4h\right]\Bigg) \ .
 \label{Atm2}
 \end{eqnarray}
Fig. \ref{Fig2} shows a small deviation from the TBM atmospheric mixing angle with $\theta_{23}>45^\circ$,
recalling that $\theta_{23}<45^\circ$ in case (i). It is thus crucial to have precise measurements of the
atmospheric mixing angle in the future to see whether $\theta_{23}\leq45^{\circ}$ or $\theta_{23}\geq45^{\circ}$
in order to test different scenarios.
%Since there is no first order of $\lambda$ or $\epsilon$ in Eq.~(\ref{Atm2}), deviation from maximal
%mixing comes mainly from the terms associated with $\lambda^{2}$ or $\epsilon\lambda$.
For $\sin\delta\approx1$ the deviation from the maximal mixing of $\theta_{23}$ is approximated as
$\sin^{2}\theta_{23}-\frac{1}{2}\approx\frac{\epsilon\lambda}{\sqrt{3}}+\lambda^{2}(f-\frac{1}{4})$, which
leads to $\sin^2\theta_{23}>1/2$ for $0<|\epsilon|<\lambda$. The behavior of $\sin^{2}\theta_{23}$ is
plotted in Fig.~\ref{Fig2} as a function of $\delta$.
Likewise, the reactor mixing angle $\theta_{13}$ can be written as
 \begin{eqnarray}
  \sin^2\theta_{13}&=&\frac{2\epsilon^{2}}{3}+\frac{2\epsilon\lambda\sin\delta}{3}
  (\epsilon\cos\delta-\sqrt{3}\sin\delta)+\lambda^{2}\left(\frac{1}{2}-\epsilon^{2}\right) \nonumber \\
  &+& \lambda^{3}\epsilon\left(\frac{\sin\delta(1-2f)}{\sqrt{3}}-\frac{\epsilon\cos\delta(1+2f)}{3}\right)~.
 \label{Reactor2}
 \end{eqnarray}
%which indicates that $\sin\theta_{13}$ depends sizably on the parameters $\lambda$ and $\epsilon$.
%Especially, when $\epsilon$ goes to zero $\sin\theta_{13}$ approaches $\lambda/\sqrt{2}$~\cite{Ahn:2011ep}.
%Note here that the size of $\epsilon$ is constrained by the plot $\sin\theta_{13}$ versus $\delta$ in
%Fig.~\ref{Fig2} where the horizontal dot-dashed lines represent the present data of T2K Collaboration
%for normal hierarchy, see Eq.~(\ref{T2K}). For a negative value of $\epsilon$ the plot for $\sin\theta_{13}$
%versus $\delta$ is flipped $180^{\circ}$ around the axis $\delta=3.14$ [rad]. If we assume $\rho>0$, we
%see from Eq.~(\ref{masssquare}) that a positive (negative) value of $\epsilon$ gives normal (inverted)
%neutrino mass spectrum, and which in turn implies $\epsilon\leq0.12$ ($\epsilon\geq-0.07$) for $\delta=1.56$ [rad].
We find $0.07 \lesssim\sin\theta_{13}\leq\frac{\lambda}{\sqrt{2}}$
($\frac{\lambda}{\sqrt{2}}\leq\sin\theta_{13}\lesssim 0.22$) for $\delta=1.56$ and $\epsilon\leq0.12$
($\epsilon\geq-0.07$)\ .

The Dirac phase $\delta_{CP}$ has the expression
 \begin{eqnarray}
  \delta_{CP}&=&\tan^{-1}\left(\frac{\lambda\{\sqrt{3}(2-\epsilon^{2})(1-f\lambda^{2})
  \sin\delta-6\epsilon(1+f\lambda^{2})\cos\delta\}}{\sqrt{3}(2-\epsilon^{2})\{2-\lambda^{2}
  +(1-f\lambda^{2})\lambda\cos\delta\}+6\epsilon\lambda(1+f\lambda^{2})\sin\delta}\right) \nonumber\\
  &-&\tan^{-1}\left(\frac{\lambda\{2\sqrt{3}\epsilon(1-f\lambda^{2})\sin\delta+3(2-\epsilon^{2})
  (1+f\lambda^{2})\cos\delta\}}{3\lambda(\epsilon^{2} -2)(1+f\lambda^{2})\sin\delta+2\sqrt{3}\epsilon(2-\lambda^{2}+(1-f\lambda^{2})\cos\delta)}\right)~.
  \label{Diracphase2}
 \end{eqnarray}
Assuming $\rho>0$, we show in Table~\ref{DiracCP21} the predictions for $\delta_{\rm CP}$ and $\theta_{13}$ as a function of $\epsilon$, where we have focused on the central values of Eq.~(\ref{eq:QMfh}).
\begin{center}
\begin{table}[h]
\caption{\label{DiracCP21} Predictions of $\delta_{CP}$ and $\theta_{13}$ as a function of $\epsilon$ in the case of $V^{\ell}_{L}=V_{\rm QM}$.}
\begin{ruledtabular}
\begin{tabular}{ccc}
 $\epsilon$ & $\delta_{CP}~[{\rm deg.}]$ & $\theta_{13}~[{\rm deg.}]$ \\
\hline
 $0\sim0.08$ & $-7.1\sim-5.2$ & $9.1\sim12.6$ \\
 $-0.11\sim0$ & $-15.6\sim-7.1$ & $4.1\sim9.1$
\end{tabular}
\end{ruledtabular}
\end{table}
\end{center}
The strength of $CP$ violation $J_{CP}$ can be expressed in a similar way  to Eq.~(\ref{JCP1})
 \begin{eqnarray}
  J_{CP}&=&-\frac{\epsilon}{3\sqrt{3}}+\frac{\lambda}{6}\left(\sin\delta+\epsilon\frac{4\cos\delta}
  {\sqrt{3}}-2\epsilon^{2}\sin\delta\right) \\
  &+&\frac{\lambda^{2}}{9}\left(\sin\delta(h-\cos\delta)+\epsilon\sqrt{3}(1-f-\cos2\delta-h\cos\delta)
  -\epsilon^{2}(h-4\cos\delta)\sin\delta\right), \nonumber
 \label{JCP2}
 \end{eqnarray}
which can be approximated as $J_{CP}\approx-\frac{\epsilon}{3\sqrt{3}}+\frac{\lambda}{6}\sin\delta$.
When $\sin\delta\approx1$, it is further reduced to
$J_{CP}\approx-\frac{\epsilon}{3\sqrt{3}}+\frac{\lambda}{6}\leq\frac{\lambda}{6}~(\geq\frac{\lambda}{6})$
for $\epsilon>0~(\epsilon<0)$. Assuming $\rho>0$, we see from Fig.~\ref{Fig2} that
$0.014\lesssim J_{CP}\lesssim0.037$ ($0.037\lesssim J_{CP}\lesssim0.05$) for $\epsilon\leq0.12$
($\epsilon\geq-0.07$) and $\delta=1.56$\ .

%%%%%%%%%%%%%%%%%%%%%%%%%%%%%%%%%%%%%%%%%%%%%%%%%%%%%%%%%%%%%%%%%%%%%%%%%%%%%%%%%%%%%%%%%%
\section{Conclusion}

In their original work, Harrison, Perkins and Scott proposed simple charged lepton and neutrino mass
matrices that lead to the tribimaximal mixing $U_{\rm TBM}$. In this paper we considered a general
extension of the mass matrices so that the lepton mixing matrix becomes
$U_{\rm PMNS}=V_L^{\ell\dagger}U_{\rm TBM}W$.
Hence, corrections to  the tribimaximal mixing arise from both charged lepton and neutrino sectors: the
charged lepton mixing matrix $V_L^\ell$ measures the deviation of  from the trimaximal form and the $W$
matrix characterizes the departure of the neutrino mixing from the bimaximal one.
Following our previous work to assume a Qin-Ma-like parametrization $V_{\rm QM}$ for $V_L^\ell$ in which
the {\it CP}-odd phase is approximately maximal, we study the phenomenological implications in two
different scenarios: $V_L^\ell=V_{\rm QM}^\dagger$ and $V_L^\ell=V_{\rm QM}$.
We found that both scenarios are consistent with the data within $3\sigma$ ranges. Especially, the
predicted central value of the reactor neutrino mixing angle $\theta_{13}=9.2^\circ$ is in good agreement
with the recent T2K data.
However, the data of $\sin^2\theta_{12}$ can be easily accommodated in the second scenario but only
marginally in the first one. Hence, the precise measurements of the solar mixing angle in future
experiments will test which scenario is more preferable.  The leptonic {\it CP} violation characterized by
the Jarlskog invariant $J_{\rm CP}$ is generally of order $10^{-2}$.

%%%%%%%%%%%%%%%%%%%%%%%%%%%%%%%%%%%%%%%%%%%%%%%%%%%%%%%%%%%%%%%%%%%%%%%%%%%%%%%%%%%%%%%%%%%%%%%%%%%%%%%%%%%%%%%%
\vskip 1 cm {\bf Acknowledgments}

This work was supported in part by the National Science Council of R.O.C. under Grants Numbers:  NSC-97-2112-M-008-002-MY3, NSC-100-2112-M-001-009-MY3 and NSC-99-2811-M-001-038.

%%%%%%%%%%%%%%%%%%%%%%%%%%%%%%%%%%%%%%%%%%%%%%%%%%%%%%%%%%%%%%%%%%%%%%%%%%%%%%%%%%%%%%%%%%%%%%%%%%%%%%%%%%%
%%%%%%%%%%%%%%%%%%%%%%%%%%%%%%%%%%%%%%%%%%%%%%%%%%%%%%%%%%%%%%%%%%%%%%%%%%%%%%%%%%%%%%%%%%%%%%%%%%%%%%%%%%%


\begin{thebibliography}{99}
\def\plb#1#2#3{Phys.\ Lett.\       {\bf B#1}, (#3) #2}
\def\npb#1#2#3{Nucl.\ Phys.\       {\bf B#1}, (#3) #2}
\def\prd#1#2#3{Phys.\ Rev.\        {\bf D#1}, (#3) #2}
\def\prl#1#2#3{Phys.\ Rev.\ Lett.\ {\bf #1},  (#3) #2}
\def\mpl#1#2#3{Mod.\ Phys.\ Lett.\ {\bf A#1}, (#3) #2}
\def\rep#1#2#3{Phys.\ Rep.\        {\bf #1},  (#3) #2}
\def\sci#1#2#3{Science             {\bf #1},  (#3) #2}
\def\astro#1#2#3{Astrophys.\ J.\   {\bf #1},  (#3) #2}
\def\epj#1#2#3{Eur.\ Phys.\ J.  {\bf C#1},  (#3) #2}
\def\jhep#1#2#3{JHEP               {\bf #1},  (#3) #2}
\def\jpg#1#2#3{J.\ Phys.\        {\bf G#1},  (#3) #2}
\def\ijmp#1#2#3{Int.\ J.\ Mod.\ Phys.\ {\bf #1},  (#3) #2}
\def\ptp#1#2#3{Prog.\ Theor.\ Phys.\ {\bf #1},  (#3) #2}
%%%%%%%%%%%%%%%%%%%%%%%%%%%%%%%%%%%%%%%%%%%%%%%%%%%%%%%%%%%%%%
\bibitem{HPS}
  P.~F.~Harrison, D.~H.~Perkins and W.~G.~Scott,
  %``Tri-bimaximal mixing and the neutrino oscillation data,''
  Phys.\ Lett.\  B {\bf 530}, 167 (2002)
  [arXiv:hep-ph/0202074];
  P.~F.~Harrison and W.~G.~Scott,
  %``Symmetries and generalizations of tri - bimaximal neutrino mixing,''
  Phys.\ Lett.\  B {\bf 535}, 163 (2002)
  [arXiv:hep-ph/0203209].
  %%CITATION = PHLTA,B535,163;%%

\bibitem{He:2006dk}
  X.~G.~He, Y.~Y.~Keum and R.~R.~Volkas,
  %``A(4) flavor symmetry breaking scheme for understanding quark and neutrino
  %mixing angles,''
  JHEP {\bf 0604}, 039 (2006)
  [arXiv:hep-ph/0601001].
  %%CITATION = JHEPA,0604,039;%%


\bibitem{Abe:2011sj}
  K.~Abe {\it et al.}  [T2K Collaboration],
  %``Indication of Electron Neutrino Appearance from an Accelerator-produced
  %Off-axis Muon Neutrino Beam,''
  Phys. Rev. Lett. {\bf 107}, 041801 (2011)
  [arXiv:1106.2822 [hep-ex]].
  %%CITATION = ARXIV:1106.2822;%%

\bibitem{MINOS}
P. Adamson {\it et al.} [MINOS Collaboration], arXiv:1108.0015 [hep-ex].

\bibitem{GonzalezGarcia:2010er}
  M.~C.~Gonzalez-Garcia, M.~Maltoni and J.~Salvado,
  %``Updated global fit to three neutrino mixing: status of the hints of theta13
  %> 0,''
  JHEP {\bf 1004}, 056 (2010)
  [arXiv:1001.4524v3 [hep-ph]].
  %%CITATION = JHEPA,1004,056;%%

\bibitem{Fogli:2011qn}
  G.~L.~Fogli, E.~Lisi, A.~Marrone, A.~Palazzo and A.~M.~Rotunno,
  %``Evidence of $\theta_{13}$>0 from global neutrino data analysis,''
  Phys.\ Rev.\  D {\bf 84}, 053007 (2011)
  [arXiv:1106.6028 [hep-ph]].
  %%CITATION = PHRVA,D84,053007;%%

\bibitem{PDG}
 K.~Nakamura {\it et al}. (Particle Data Group), J. Phys. G {\bf 37}, 075021 (2010).

\bibitem{Schechter:1981bd}
  J.~Schechter and J.~W.~F.~Valle,
  %``Neutrinoless Double beta Decay in SU(2) x U(1) Theories,''
  Phys.\ Rev.\  D {\bf 25}, 2951 (1982).
  %%CITATION = PHRVA,D25,2951;%%

\bibitem{Cabibbo:1977nk}
  N.~Cabibbo,
  %``Time Reversal Violation in Neutrino Oscillation,''
  Phys.\ Lett.\  B {\bf 72}, 333 (1978).
  %%CITATION = PHLTA,B72,333;%%


\bibitem{trimax}
L. Wolfenstein, Phys. Rev. D {\bf 18}, 958 (1978);
P. F. Harrison and W. G. Scott,  Phys. Lett. B {\bf 333}, 471 (1994);
R. N. Mohapatra and S. Nussinov, Phys. Lett. B {\bf 346}, 75 (1995).

\bibitem{Xing:2002sw}
  Z.~Z.~Xing,
  %``Nearly tri bimaximal neutrino mixing and CP violation,''
  Phys.\ Lett.\  B {\bf 533}, 85 (2002)
  [arXiv:hep-ph/0204049].


\bibitem{TBMnu}
  Y.~Shimizu, M.~Tanimoto and A.~Watanabe,
  %``Breaking Tri-bimaximal Mixing and Large $\theta_{13}$,''
  Prog.\ Theor.\ Phys.\  {\bf 126}, 81 (2011)
  [arXiv:1105.2929 [hep-ph]];
   N.~Qin and B.~Q.~Ma,
  %``A New relation between quark and lepton mixing matrices,''
  Phys.\ Lett.\  B {\bf 702}, 143 (2011)
  [arXiv:1106.3284 [hep-ph]];
   Y.~J.~Zheng and B.~Q.~Ma,
  %``Re-evaluation of neutrino mixing pattern according to latest T2K result,''
  arXiv:1106.4040 [hep-ph];
  E.~Ma and D.~Wegman,
  %``Nonzero theta(13) for neutrino mixing in the context of A(4) symmetry,''
  Phys.\ Rev.\ Lett.\  {\bf 107}, 061803 (2011)
  [arXiv:1106.4269 [hep-ph]];
  X.~G.~He and A.~Zee,
  %``Minimal Modification to Tri-bimaximal Mixing,''
  Phys.\ Rev.\  D {\bf 84}, 053004 (2011)
  [arXiv:1106.4359 [hep-ph]];
    T.~Araki,
  %``Getting at large $\theta_{13}$ with almost maximal $\theta_{23}$ from
  %tri-bimaximal mixing,''
  Phys.\ Rev.\  D {\bf 84}, 037301 (2011)
  [arXiv:1106.5211 [hep-ph]];
   S.~Morisi, K.~M.~Patel and E.~Peinado,
  %``Model for T2K indication with maximal atmospheric angle and tri-maximal
  %solar angle,''
  Phys.\ Rev.\  D {\bf 84}, 053002 (2011)
  [arXiv:1107.0696 [hep-ph]];
   W.~Chao and Y.~J.~Zheng,
  %``Relatively Large Theta13 from Modification to the Tri-bimaximal, Bimaximal
  %and Democratic Neutrino Mixing Matrices,''
  arXiv:1107.0738 [hep-ph];
  S.~Dev, S.~Gupta and R.~R.~Gautam,
  %``Parametrizing the Lepton Mixing Matrix in terms of Charged Lepton
  %Corrections,''
  Phys.\ Lett.\  B {\bf 704}, 527 (2011)
  [arXiv:1107.1125 [hep-ph]];
  R.~d.~A.~Toorop, F.~Feruglio and C.~Hagedorn,
  %``Discrete Flavour Symmetries in Light of T2K,''
  Phys.\ Lett.\  B {\bf 703}, 447 (2011)
  [arXiv:1107.3486 [hep-ph]].



\bibitem{Ahn:2011ep}
  Y.~H.~Ahn, H.~Y.~Cheng and S.~Oh,
  %``Towards a realistic tribimaximal-like neutrino mixing matrix,''
  arXiv:1105.4460 [hep-ph] (unpublished).
  %%CITATION = ARXIV:1105.4460;%%


\bibitem{TBMlep}
  S.~Dev, S.~Gupta and R.~R.~Gautam,
  %``Parametrizing the Lepton Mixing Matrix in terms of Charged Lepton
  %Corrections,''
  Phys.\ Lett.\  B {\bf 704}, 527 (2011)
  [arXiv:1107.1125 [hep-ph]];
  P.~S.~Bhupal Dev, R.~N.~Mohapatra and M.~Severson,
  %``Neutrino Mixings in SO(10) with Type II Seesaw and $\theta_{13}$,''
  Phys.\ Rev.\  D {\bf 84}, 053005 (2011)
  [arXiv:1107.2378 [hep-ph]].

\bibitem{Ahn:2011yj}
  Y.~H.~Ahn, H.~Y.~Cheng and S.~Oh,
  %``Quark-lepton complementarity and tribimaximal neutrino mixing from discrete
  %symmetry,''
  Phys.\ Rev.\  D {\bf 83}, 076012 (2011)
  [arXiv:1102.0879 [hep-ph]].
  %%CITATION = PHRVA,D83,076012;%%


\bibitem{CKM}
N. Cabibbo, Phys. Rev. Lett. {\bf 10}, 531 (1963);
  M.~Kobayashi and T.~Maskawa,
  %``CP Violation in the Renormalizable Theory of Weak Interaction,''
   Prog.\ Theor.\ Phys.\  {\bf 49}, 652 (1973).


\bibitem{Qin:2011ub}
  N.~Qin and B.~Q.~Ma,
  %``Parametrization of fermion mixing matrices in Kobayashi-Maskawa form,''
  Phys.\ Rev.\  D {\bf 83}, 033006 (2011)
  [arXiv:1101.4729 [hep-ph]].
  %%CITATION = PHRVA,D83,033006;%%

\bibitem{Wolfenstein:1983yz}
  L.~Wolfenstein,
  %``Parametrization Of The Kobayashi-Maskawa Matrix,''
  Phys.\ Rev.\ Lett.\  {\bf 51}, 1945 (1983).
  %%CITATION = PRLTA,51,1945;%%

\bibitem{Koide}
  Y.~Koide and H.~Nishiura,
  %``Maximal CP Violation Hypothesis and a Lepton Mixing Matrix,''
  Phys.\ Rev.\  D {\bf 79}, 093005 (2009)
  [arXiv:0811.2839 [hep-ph]].


\bibitem{CKMfitter}
CKMfitter Group, J. Charles {\it et al.,} Eur.
Phys. J. C {\bf 41}, 1 (2005) and updated results from
http://ckmfitter.in2p3.fr.


%\cite{Ahn:2011it}
\bibitem{Ahn:2011it}
  Y.~H.~Ahn, H.~Y.~Cheng, S.~Oh,
  %``Remarks on the Qin-Ma Parametrization of Quark Mixing Matrix,''
  Phys.\ Lett.\ B {\bf701}, 614 (2011)
  [arXiv:1105.0450 [hep-ph]].
  %%CITATION = ARXIV:1105.0450;%%



\bibitem{Branco:2002xf}
  G.~C.~Branco, R.~Gonzalez Felipe, F.~R.~Joaquim, I.~Masina, M.~N.~Rebelo and C.~A.~Savoy,
  %``Minimal scenarios for leptogenesis and CP violation,''
  Phys.\ Rev.\  D {\bf 67}, 073025 (2003)
  [arXiv:hep-ph/0211001].
  %%CITATION = PHRVA,D67,073025;%%


\bibitem{Jarlskog:1985ht}
  C.~Jarlskog,
  %``Commutator of the Quark Mass Matrices in the Standard Electroweak Model and
  %a Measure of Maximal CP Violation,''
  Phys.\ Rev.\ Lett.\  {\bf 55}, 1039 (1985);
  %%CITATION = PRLTA,55,1039;%%
  D.~d.~Wu,
  %``The Rephasing Invariants and CP,''
  Phys.\ Rev.\  D {\bf 33}, 860 (1986).
  %%CITATION = PHRVA,D33,860;%%





%%%%%%%%%%%%%%%%%%%%%%%%%%%%%%%%%%%%%%%%%%%%%%%%%%%%%%%%%%%%%%%%%%%%%%%%%%%%%%%%%%%%%%
\end{thebibliography}
\end{document}